\documentclass[twocolumn, prl, superscriptaddress]{revtex4-2}
\usepackage{graphicx}
\usepackage{physics}
\usepackage{color}
\usepackage{amsmath}
\usepackage{amsfonts}
\usepackage{xspace}
\usepackage{array}
\usepackage{makecell}
\usepackage{ulem}
\usepackage[dvipsnames]{xcolor}
\usepackage[breaklinks,colorlinks,linkcolor=blue,citecolor=blue,urlcolor=blue]{hyperref}
\newcommand{\di}{{\rm{d}}}
\newcommand{\Poincare}{Poincar\'e\xspace}

\DeclareMathOperator\arcsinh{arcsinh}

\begin{document}

\title{Emergent spacetimes from Hermitian and non-Hermitian quantum dynamics}
\author{Chenwei Lv}
\affiliation{Department of Physics and Astronomy, Purdue University, West Lafayette, IN, 47907, USA}\author{Qi Zhou}
\email{zhou753@purdue.edu}
\affiliation{Department of Physics and Astronomy, Purdue University, West Lafayette, IN, 47907, USA}
\affiliation{Purdue Quantum Science and Engineering Institute, Purdue University, West Lafayette, IN 47907, USA}
\date{\today}
\begin{abstract}

We show that quantum dynamics of any systems with $SU(1,1)$ symmetry give rise to emergent Anti-de Sitter spacetimes in 2+1 dimensions (AdS$_{2+1}$). 
Using the continuous circuit depth, a quantum evolution is mapped to a trajectory in AdS$_{2+1}$. Whereas the time measured in laboratories becomes either the proper time or the proper distance, quench dynamics follow geodesics of AdS$_{2+1}$. Such a geometric approach provides a unified interpretation of a wide range of prototypical phenomena that appear disconnected. 
For instance, the light cone of AdS$_{2+1}$ underlies expansions of unitary fermions released from harmonic traps, the onsite of parametric amplifications, and the exceptional points that represent the $PT$ symmetry breaking in non-Hermitian systems. 
Our work provides a transparent means to optimize quantum controls by exploiting shortest paths in the emergent spacetimes. It also allows experimentalists to engineer emergent spacetimes and induce tunnelings between different AdS$_{2+1}$.

\end{abstract}
\maketitle 

Studies have shown that spacetimes could arise as emergent phenomena in quantum systems. In certain strongly coupled systems, the underlying gauge theory has a gravitational counterpart with an extra dimension~\cite{Maldacena1998,Gubser1998,Witten1998,Vidal2007,Brian2012,Nozaki2012,Qi2013,Preskill2015}. It has also been found that circuit depth, a fundamental concept in quantum computation that represents the number of steps to reach a target state, can be visualized using certain geometries, leading to synthetic spacetimes in quantum dynamics~\cite{Nielsen2005,Nielsen2006,Jefferson2017,Chapman2018,Guo2018,Susskind2019}. 
These emergent spacetimes provide physicists a new means to connect condensed matter physics and quantum information to high energy physics.  

To derive an emergent spacetime metric from quantum dynamics, a continuous version of circuit depth could be used~\cite{Nielsen2005,Nielsen2006,Jefferson2017,Chapman2018,Guo2018,Susskind2019}. 
Consider quantum gates that are produced from a collection of unitary operators generated by $\{K_i\}$, a generic gate can be expressed as $e^{-i \sum_i Y_iK_i}$, where $Y_i$ are real numbers. 
A sequence of gates turns a reference state $|R\rangle$ prepared at the initial time $\tau=0$ into a target state at time $\tau$, $|T\rangle= U|R\rangle$, where $U={\text T} e^{-i \int_0^\tau d\tau' [\sum_i \xi_i(\tau')K_i]/\hbar}$ is the propagator and ${\text T}$ is the time-ordering operator. 
In a single gate operated in a time interval between $\tau'$ and $\tau'+d\tau'$, $Y_i=\xi_i(\tau') d\tau'/\hbar$, where $\xi_i$ is the $i$th component of a field coupled to $K_i$. 
The line element in the geometry is written as~\cite{Nielsen2005}
\begin{equation}
ds= F[Y_i(\tau')],
\end{equation}
where $F[Y_i]$ is a chosen function based on certain physical considerations. 
For instance, if $F[Y_i]=\sum_i|Y_i|$, $S=\int ds$ amounts to the total number of steps of all operations produced by independent $K_i$ to reach the target state. 
A variety of other functions were considered in the literature~\cite{Nielsen2005,Jefferson2017,Susskind2019,Sood2022}. 

Here, we consider generators of the $SU$(1,1) group,
\begin{equation}
  [K_0,K_1]=iK_2,\, [K_1,K_2]=-iK_0,\,[K_2,K_0]=iK_1,
\end{equation}
which are fundamental ingredients in holographic tensor networks, determining how quantum entanglement is developed with changing the length scales~\cite{Nozaki2012,Qi2013}. 
They also produce Hamiltonians of a wide range of quantum systems with $SU$(1,1) symmetry, i.e., $H(\tau)=\sum_i\xi_i(\tau) K_i$, where $\xi_i$ are real~\cite{echo1,echo2,echo3,Cheng2021,Zhang2022}. 
Table~\ref{table1} shows a few examples in few-body and many-body systems, which correspond to different representations of the $SU$(1,1) group. 

We summarize our main results as follows. Our scheme unifies a wide range of prototypical topics in quantum physics, including but not limited to one mode squeezing, two mode squeezing, Efimov states, scale-invariant quantum systems, and spins in complex magnetic fields. 
Despite that these topics were studied in distinct content, we show that they are governed by the same underlying spacetimes, an AdS$_{2+1}$. 
In particular, the exceptional point in non-Hermitian quantum systems and the parametric amplification in Hermitian quantum systems are different sides of the same coin, being the light cone of AdS$_{2+1}$. 
Our results provide experimentalists an efficient means to reach any desired target states with the least time by using the geodesics in the geometry. 
Furthermore, our work allows experimentalists to bypass constraints from Einstein equation and create intriguing synthetic spacetimes in laboratories.

We consider 
\begin{equation}
  F[Y_i]= \sqrt{-Y_0^2+Y_1^2+Y_2^2}/2\label{cost},
\end{equation}
which is motivated by the following observations. 
Consider a Hamiltonian $H(\tau)=\sum_i\xi_i(\tau) K_i$, the instantaneous eigenenergies are written as $\pm (m+k)\sqrt{\xi_0^2-\xi_1^2-\xi_2^2}$, where $m$ is an integer, $k$ is the Bargmann index.
The strength of the field coupled to $\{K_i\}$ is thus $\xi=\sqrt{\xi_0^2-\xi_1^2-\xi_2^2}$, as analogous to the strength of a magnetic field coupled to a spin in the $su$(2) case. 
The equally spaced instantaneous eigenenergies thus set a natural time scale $\sim \hbar/\xi$. 
While it certainly takes less time to reach a target state by applying a larger external field, there are always constraints on the largest possible external fields accessible in realistic experiments. 
As such, it is desired to consider the shortest possible time at a constant $\xi$. The line element,
\begin{equation}
ds=\sqrt{-\xi_0^2+\xi_1^2+\xi_2^2} d\tau/(2\hbar) \label{line}  
\end{equation}
is equivalent to the experimental time $d\tau$ once the field strength $\xi$ is fixed. 
We thus directly correlate the experimental time to circuit depth and obtain the shortest possible time to achieve the target state using the information provided by the metric of the emergent spacetime. 
$F$ defined in Eq.(\ref{cost}) can be either real or imaginary, corresponding to whether it is spacelike or timelike in the emergent spacetime. 

\begin{figure}[h] 
  \includegraphics [width=0.5\textwidth]
  {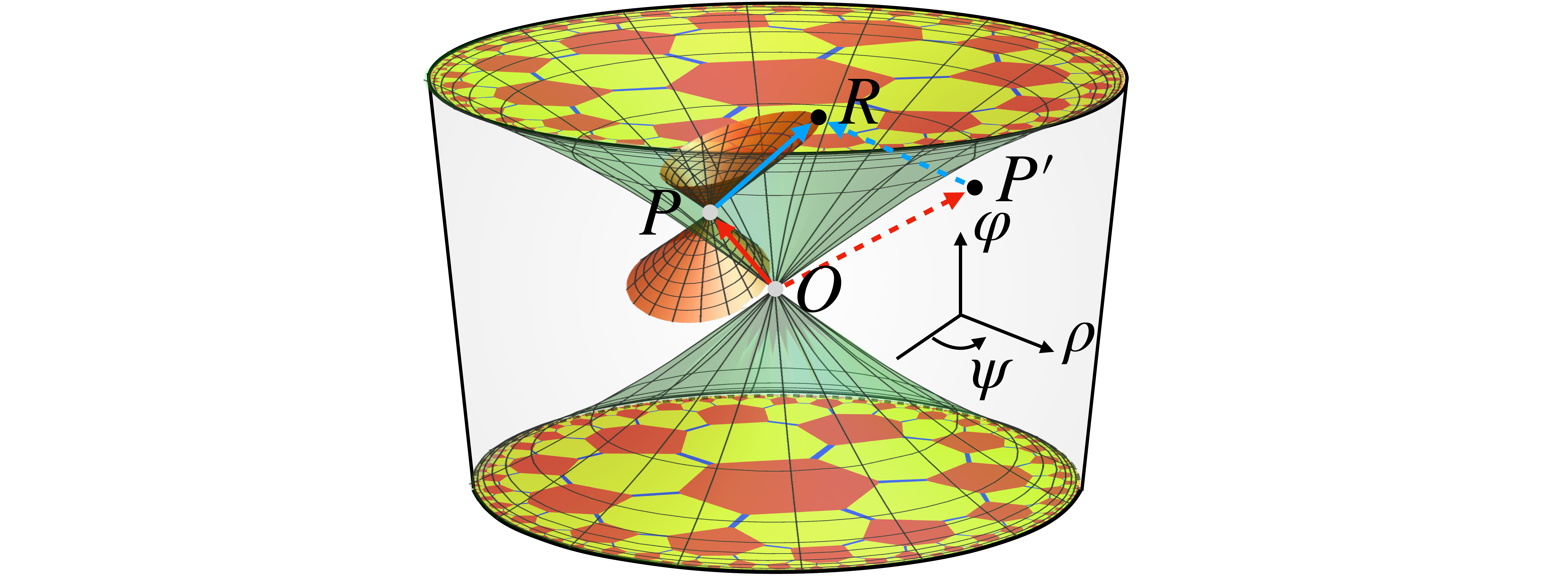}
  \caption
  {
  An AdS$_{2+1}$. 
  Each cross section with a constant $\varphi$ is a \Poincare disk. 
  Starting from the initial state $O$, the shortest path $OPR$ minimizes the proper length or equivalently, the time spent from $O$ to $R$. 
  $OP$ and $PR$ lay on the light cone of $O$ and $P$, respectively. 
  A deviation from this trajectory, for instance $OP'R$, where $P'$ is not on the light cone of $O$, increases the time to reach the same target state $R$. 
  }
  \label{fig:fig1}
\end{figure}

We use a parameterization of the propagator $U$~\cite{Puri2001}, 
\begin{equation}
U=e^{-iK_0(\varphi-\psi)}e^{-2iK_1\eta}e^{-iK_0(\varphi+\psi)}\label{PG},
\end{equation}
where $\eta$, $\psi$ and $\varphi$ are functions of $\tau$. 
Since $i\hbar\frac{dU}{d\tau}U^{-1}=\sum_{i=0}^2\xi_i(\tau) K_i$, we obtain $\xi_i={\text Tr}[i\hbar\frac{dU}{d\tau}U^{-1}K_i^\dag]$, with Cartan-Killing inner product ${\text Tr}(K_iK_j^\dag)=\delta_{ij}$~\cite{Gilmore2008}.
Substituting Eq.(\ref{PG}) to the right-hand side of this expression, $\xi_i$ is expressed as a function of $\eta$, $\psi$, $\varphi$ and $d\eta/d\tau$, $d\psi/d\tau$, $d\varphi/d\tau$. 
The line element in Eq.(\ref{line}) is then expressed in terms of $\eta$, $\psi$, $\varphi$ and $d\eta$, $d\psi$, $d\varphi$,
\begin{equation}
  ds^2=\di\eta^2-\cosh^2(\eta)\di\varphi^2+\sinh^2(\eta)\di\psi^2\label{eq:ads}.
\end{equation}
Eq.\ref{eq:ads} describes a curved spacetime with a constant negative curvature, known as the Anti-de Sitter (AdS) spacetime in (2+1) dimensions. 
We emphasize that this spacetime needs to be considered as a synthetic one, since it does not arise from the Einstein equation. 
Whereas an AdS in gravity requires a negative cosmological constant~\cite{Bengtsson2006}, here, it emerges from the symmetry group governing the quantum dynamics. 
A coordinate transformation $\eta=2{\rm arctanh}(\rho)$ leads to 
\begin{equation}
  ds^2=-\left(\frac{1+\rho^2}{1-\rho^2}\right)^2\di\varphi^2+\frac{4}{(1-\rho^2)^2}\left(\di\rho^2+\rho^2\di\psi^2\right)\label{eq:ads_s},
\end{equation}
and brings the boundary $\eta=\infty$ to $\rho= 1$ as shown in Fig.~\ref{fig:fig1}.
Starting from a reference state, which is placed at the origin, any quantum evolution governed by a Hamiltonian $H=\sum_{i=0}^2\xi_iK_i$ is mapped to a trajectory $(\eta(\tau),\varphi(\tau),\psi(\tau))$ in this AdS$_{2+1}$. 
When $\varphi$ is constant, $ds^2=d\eta^2+\sinh^2(\eta)d\psi^2$, AdS$_{2+1}$ reduces to a \Poincare disk, recovering the geometry we previously obtained about geometrizing $SU(1,1)/U(1)$~\cite{echo2}. 
Whereas we have been focusing on isotropic and homogeneous $F$, our results can be generalized to anisotropic or non-uniform $F$ that may produce even more intriguing spacetimes (Supplemental Materials). 

%-------------------------------------------------------------------------
%-------------------------------------------------------------------------

If $\xi=0$, from Eq.(\ref{line}), we see that this gives rise to $ds=0$, i.e., the light cone in AdS$_{2+1}$. 
Using Eq.(\ref{eq:ads}), the explicit expression of the light cone can be obtained from $d\eta^2=\cosh^2(\eta)d\varphi^2-\sinh^2(\eta)d\psi^2$.  
The light cone centered at the origin is written as
\begin{equation}
  \varphi\pm {\rm arctan}(\sinh(\eta))=0.
\end{equation}
When $\xi^2<0$ ($\xi^2>0$), the quantum evolution is mapped to a trajectory in the spacelike (timelike) regime. 

Being the local extreme of the proper time or the proper distance, geodesics distinguish themselves from other trajectories.
Counterparts of geodesics in AdS$_{2+1}$ turn out to be quantum dynamics with time-independent $\xi_{i=0,1,2}$, $\Tr[\frac{d}{d\tau}\left(i\frac{dU}{d\tau}U^{-1}\right)K^\dag_i]=0$. 
Applying the parameterization of $U$ in Eq.~\ref{PG}, we obtain
\begin{equation}
    \begin{split}
    &\frac{d^2\varphi}{d\tau^2}=-2\tanh(\eta)\frac{d\varphi}{d\tau}\frac{d\eta}{d\tau},\\
    &\frac{d^2\eta}{d\tau^2}=-\sinh(\eta)\cosh(\eta)\left[\left(\frac{d\varphi}{d\tau}\right)^2-\left(\frac{d\psi}{d\tau}\right)^2\right],\\
    &\frac{d^2\psi}{d\tau^2}=-2\coth(\eta)\frac{d\psi}{d\tau}\frac{d\eta}{d\tau}.
  \end{split}\label{eq:geo}
\end{equation}
These are geodesic equations in AdS$_{2+1}$ (Supplemental Materials). 

%-------------------------------------------------------------------------
\begin{table*}[t]
  \centering
  \resizebox{2.\columnwidth}{!}{
  \begin{tabular}{|>
    {\centering\arraybackslash}m{5.4cm}|>
    {\centering\arraybackslash}m{5.6cm}|>
    {\centering\arraybackslash}m{5.6cm}|>
    {\centering\arraybackslash}m{3.8cm}|}
  \hline 
  \makecell{} &  
  \makecell{ $K_0$ } & 
  \makecell{ $K_1$ } & 
  \makecell{ $K_2$ }\\
  \hline 
  \makecell{One mode squeezing} & 
  \makecell{$(2a^\dag a+1)/4$} & 
  \makecell{$(a^{\dag 2}+a^2)/4$}&
  \makecell{$(a^{\dag 2}-a^2)/(4i)$}\\
  \hline
  \makecell{Two mode squeezing} & 
  \makecell{$(a^\dag a+b^\dag b+1)/2$} & 
  \makecell{$(a^\dag b^\dag+ab)/2$}&
  \makecell{$(a^\dag b^\dag-ab)/(2i)$ }\\
  \hline
  \makecell{Efimov states} & 
  \makecell{$\big(-\partial_R^2+R^2-(1/4+s_0^2)/R^2\big)/4$} & 
  \makecell{$\big(-\partial_R^2-R^2-(1/4+s_0^2)/R^2\big)/4$} &
  \makecell{$(R\partial_R+1/2)/(2i)$}\\
  \hline
  \makecell{Scale-invariant quantum systems} & 
  \makecell{$\big(\sum_i(-\nabla_i^2+r_i^2)+\sum_{i\neq j} U(\vec{r}_i-\vec{r}_j)\big)/4$} & 
  \makecell{$\big(\sum_i(-\nabla_i^2-r_i^2)+\sum_{i\neq j} U(\vec{r}_i-\vec{r}_j)\big)/4$}&
  \makecell{$\sum_i\left(\vec{r}_i\cdot\nabla_i+\nabla_i\cdot\vec{r}_i\right)/(4i)$ }\\
  \hline
  \makecell{a spin in a complex magnetic field} & 
  \makecell{$S_0$} & 
  \makecell{$iS_1$}&
  \makecell{$iS_2$}\\
  \hline
  \end{tabular}
  }
  \caption
  {
  Generators of $su$(1,1) algebra in some Hermitian and non-Hermitian systems.
  $a$, $b$ ($a^{(\dag)}$, $b^{(\dag)}$) are bosonic annihilation (creation) operators. 
  $R$ is the hyper-radius.
  $s_0$ is the three-body parameter. 
  $R$ and $\vec{r}_i$ has the unit of harmonic length $l_0=\sqrt{\hbar/(M(0)\omega(0))}$. 
  $U$ has unit $\hbar\omega(0)$. 
  The two-body interaction satisfies $U(\lambda\vec{r})=\lambda^{-2} U(\vec{r})$ for any $\lambda$.   
  $S_{i=0,1,2}$  are angular momentum operators. 
  }\label{table1}
\end{table*}
%-------------------------------------------------------------------------

Whereas our results can be applied to any systems with $SU$(1,1) symmetry, in the simplest representation using a single particle,  $K_0$ ($K_1$) is the Hamiltonian in a harmonic (inverted harmonic) trap and $K_2$ is the scaling operator. 
More interesting representations arise in Efimov states in three-body problems and interacting many-body systems. 
For instance, the Hamiltonian of unitary fermions in a 3D harmonic trap is written as 
\begin{equation}
  H=\sum_i\left(-\frac{\hbar^2}{2M}\nabla_i^2+\frac1{2}M\omega^2(\tau) r_i^2\right)+\sum_{i<j}U(\vec{r}_i-\vec{r}_j), \label{su11h}
\end{equation}
where $U(\vec{r})$ is the interaction giving rise to a divergent scattering length. 
Without loss of generality, we have considered a time-dependent trapping frequency. 
Using the fourth row of Table~\ref{table1}, we see that
\begin{equation}
  H=\hbar\omega(\tau)\left[\left(\delta+\delta^{-1}\right)K_0-\left(\delta-\delta^{-1}\right)K_1\right],
\end{equation}
where $\delta(\tau)=\frac{M(\tau)\omega(\tau)}{M(0)\omega(0)}$. 
$\omega_0=\omega(0)$ sets the time scale of the system. 
We have defined the $SU$(1,1) generators using the Hamiltonian at $\tau=0$.
The mass $M$ is fixed for unitary fermions in 3D harmonic traps, but can, in general, be tuned as a function of time. 
For instance, the effective mass at band bottom in an optical lattice could be tuned through a time-dependent tunnelling strength~\cite{Zhang2022}.
Since $\xi_0=\hbar\omega(\delta^{-1}+\delta)$, $\xi_1=\hbar\omega(\delta^{-1}-\delta)$, when $\omega(\tau)=0$, $\xi_0=\xi_1$, and $\xi=0$. 
Turning off the harmonic trap, the free expansion of unitary fermions thus corresponds to light propagating on the light cone of AdS$_{2+1}$.
If the system is initially prepared at the many-body ground state in a harmonic trap of a frequency $\omega_0$, the trajectory in AdS$_{2+1}$ is written as $(\eta(\tau),\varphi(\tau),\psi(\tau))=(\arcsinh(\omega_0\tau/2),\arctan(\omega_0\tau/2),0)$.
Correspondingly,  $\langle K_0\rangle(\tau)=(1+\omega_0^2\tau^2/2)\langle K_0\rangle(0)$, $\langle K_1\rangle(\tau)=-\omega_0^2\tau^2/2\langle K_0\rangle(0)$.  
The radius of the fermion cloud is then written as $\langle r^2\rangle(\tau)/\langle r^2\rangle(0)=1+\omega_0^2\tau^2$.

Similarly, dynamics in a confining (repelling) potential with a positive (negative) $\omega^2$ is mapped to trajectories in the timelike (spacetime) regime. 
The repelling potential, i.e., an inverted harmonic trap, has been used to explore scrambling and quantum chaos~\cite{Morita2019,Tian2022}.
Meanwhile, unitary fermions is only one example of scale-invariant many-body systems. 
Other examples include 2D fermions and bosons with contact interactions, Calogero-Sutherland gases with $1/r^2$ interaction~\cite{Pitaevskii1997,Dalibard2019,Calogero1971,Sutherland1971,Campo2016}. 
The same conclusions also apply to these systems.

Two-mode and one-mode squeezing in quantum optics provide alternative representations of $SU$(1,1) group, as shown in Table~\ref{table1}~\cite{Yurke1986}. 
Bose-Einstein condensates with time-dependent interactions, spinor condensates with time-dependent quadratic Zeeman splitting, and fast rotating gases in the lowest Landau levels also allow experimentalists to access two-mode squeezing~\cite{Linnemann2016,Hu2019,Choi2021,Guan2021,Mukherjee2022}. 
In all these systems, parametric amplifications, which are also referred to as dynamical instability~\cite{Donley2001,Parker2013,Nguyen2017,Bloch2020,Hung2020}, may occur when $\xi^2<0$. 
As such, the light cone in the emergent AdS$_{2+1}$ underlies the onsite of parametric amplification.  
The dynamically stable (unstable) regime then corresponds to the timelike (spacelike) regime.  
Similar conclusions hold in Floquet deformed conformal field theory, where the non-heating (heating) phase corresponds to the timelike (spacelike) regime~\cite{Lapierre2020,Moosavi2021,Wen2021}.

Among all representations of $SU$(1,1) group, it is worth mentioning the non-Hermitian one, as shown by the last row of Table~\ref{table1}. 
Since the angular momentum operators $S_{i=0,1,2}$ satisfy $[S_i,S_j]=i \epsilon_{ijk}S_k$, where $\epsilon_{ijk}$ is the Levi-Civita symbol, it is clear that $K_{1,2}=iS_{1,2}$ and $K_0=S_0$ transfer the $su$(2) algebra to the $su$(1,1) algebra. 
In other words, the Hamiltonian of a spin subject to a complex magnetic field, $\vec{B}=\{B_0, B_1, B_2\}$, can be written as $H(\tau)=\sum_{i=0}^2\xi_iK_i$, or equivalently, $H(\tau)=\sum_{i=0}^2B_{i}(\tau)S_i$, where $B_0=\xi_0\in R$ and $B_{1,2}=i\xi_{1,2}\in I$. 
Such a Hamiltonian plays a fundamental role in non-Hermitian physics. 
A profound result is the existence of an exceptional point denoted by $\sum_{i=0}^2B_i^2=0$. 
Such an exceptional point signifies the $PT$ transition across which the energy spectrum becomes complex. 
It has also been widely implemented in quantum sensing~\cite{Wiersig2014,Wiersig2016,Liu2016,Hodaei2017,Chen2017,Lau2018,Zhang2019,Hokmabadi2019,Lai2019}.

A unique feature of the exceptional point is that eigenstates coalesce, making it problematic to apply the conventional method of expanding the time-dependent wavefunction using eigenstates. 
Here, dynamics at the exceptional point are dual to light propagating on the light cone of AdS$_{2+1}$. 
For instance, starting from the ground state of a Hamiltonian $H=B_0S_0$, where $B_0\in R$, a quantum quench by suddenly turning on $B_1=iB_0$, $H=B_0(S_0+iS_1)$ accesses dynamics on the light cone. 
Similarly, the $PT$ symmetry broken(unbroken) phase is mapped to spacelike(timelike) regime in AdS$_{2+1}$. 
Though circuit depth and complexity are often used in unitary evolutions of Hermitian systems, the intrinsic relation between $SU$(1,1) and $SU$(2) tells us they can also be implemented in non-Hermitian systems. 

In addition to providing a new theoretical perspective to equate quantum dynamics and geometry, the emergent AdS$_{2+1}$ offers experimentalists a unique means to manipulate quantum dynamics. 
In quantum controls, a primary question is how to fast access a target state~\cite{Torrontegui2013}. 
Since our geometric approach has transformed time spent in laboratories to length in AdS$_{2+1}$, the shortest path in such a spacetime directly unfolds the optimal choice with the least time.
Though a geodesic in AdS$_{2+1}$ is a local extreme, since AdS$_{2+1}$ is a pseudo-Riemannian manifold, the shortest length turns out to be zero. 
This can be easily seen from Fig.~\ref{fig:fig1}, which shows the shortest path corresponding to a particular two-step evolution. 
In the first step, starting from the initial state at the origin $O$, the Hamiltonian is chosen as 
$\hbar\omega_0(K_0+K_1)$ and the propagator $U_{PO}=e^{-i(K_0+K_1)\omega_0 T_1}$ produces a trajectory on the light cone of $O$. 
The length of $OP$ is thus zero. 
After reaching a particular point $P$, the Hamiltonian is quenched to $\hbar\omega_0(K_0-K_1)$ and the propagator $U_{RP}=e^{-i(K_0-K_1)\omega_0 T_2}$ in the second step connecting $P$ and $R$ falls on the light cone of $P$. 
As such, the length in the second step also vanishes. 
The total length of the shortest trajectory produced by the propagator $U_{tar}=U_{RP}U_{PO}$ is hence zero. 

%-------------------------------------------------------------------------
\begin{figure}[t] 
  \includegraphics [width=0.5\textwidth]
  {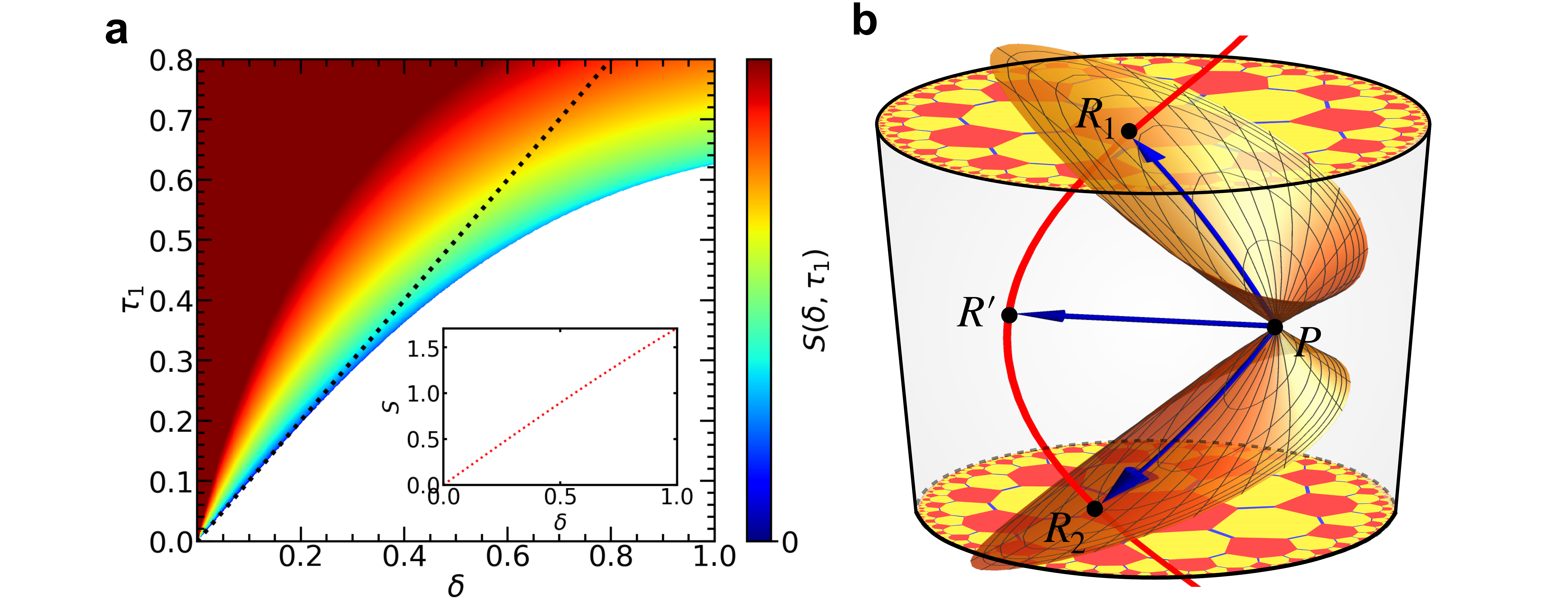}
  \caption
  {
  (a) $S(\delta,\tau_1)$ for trajectories deformed from the shortest path $OPR$.  
  $(\tau_1=T_1\delta,\delta=0)$ corresponds to $OPR$ with a vanishing $S$.
  Representing results of a particular class of deformed trajectories. 
  $T_1=1/\omega_0$, $T_2=2/\omega_0$ are used.
  (b) $PR_1$ and $PR_2$ are the shortest paths from a given initial state at $P$ to a spiral that corresponds to states with different $\varphi$. 
  It takes longer time along any other path such as $PR'$ to access the same $\eta$ and $\varphi-\psi$. 
  }
  \label{fig:fig2}
\end{figure}
%-------------------------------------------------------------------------

We also consider a deviation of the first step, i.e., $ H_1(\delta)=\hbar\omega_0\big[(\delta^{-1}-\delta)K_0+(\delta^{-1}+\delta)K_1\big]$ for $0<\tau<\tau_1$. 
We note that $e^{-iH_1(\delta)\tau_1/\hbar}$ becomes $U_{PO}$ as $\delta\to 0$ while keeping $\tau_1=T_1\delta$.
To reach the target state, the Hamiltonian in the second step is changed to $H_2=\xi_0K_0+\xi_1K_1+\xi_2K_2$ correspondingly for $\tau_1<\tau<\tau_1+\tau_2$. 
To be explicit, $H_2=\frac{i\hbar}{\tau_2}\log(U_{\rm tar}e^{iH_1(\delta)\tau_1/\hbar})$. 
The total proper length $S(\delta,\tau_1)=S_1+S_2$ can be evaluated (Supplemental Materials). 
Results depicted in Fig.~\ref{fig:fig2} clearly show that the proper length increases when $\delta\neq 0$ and $\tau_1\neq T_1\delta$. 
We therefore conclude that, for any fixed field strength $\xi$, one can adjust its direction such that the trajectory approaches the shortest path $OPR$ shown in Fig.~\ref{fig:fig1} so as to minimize the time to access the target state at $R$. 

Each point $(\eta,\psi,\varphi)$ in an AdS$_{2+1}$ can be uniquely assigned with a wavefunction. 
For instance, when considering two-mode squeezing, choosing the vacuum as the origin, $(\eta,\psi,\varphi)$ is mapped to the $SU(1,1)$ coherent state~\cite{echo2}
\begin{equation}
  \ket{\eta,\varphi,\psi}=e^{-i\varphi}\cosh^{-1}(\eta)\sum_{n=0}^\infty (-i\tanh(\eta)e^{-i(\varphi-\psi)})^n\ket{n}\label{eq:cohs}
\end{equation}
where $\ket{n}$ denotes a Fock state with $n$ particles in each bosonic mode, $\varphi$ corresponds to the overall phase, $\eta$ determines the average particle number $\bar{n}=2\sinh^2(\eta)$, and $\varphi-\psi$ gives rise to the relative phase between different Fock states. 
Whereas both $\eta$ and $\varphi-\psi$ can be measured in experiments, the overall phase $\varphi$ is more tricky. 

%-------------------------------------------------------------------------
\begin{figure}[b] 
    \includegraphics [width=0.5\textwidth]
    {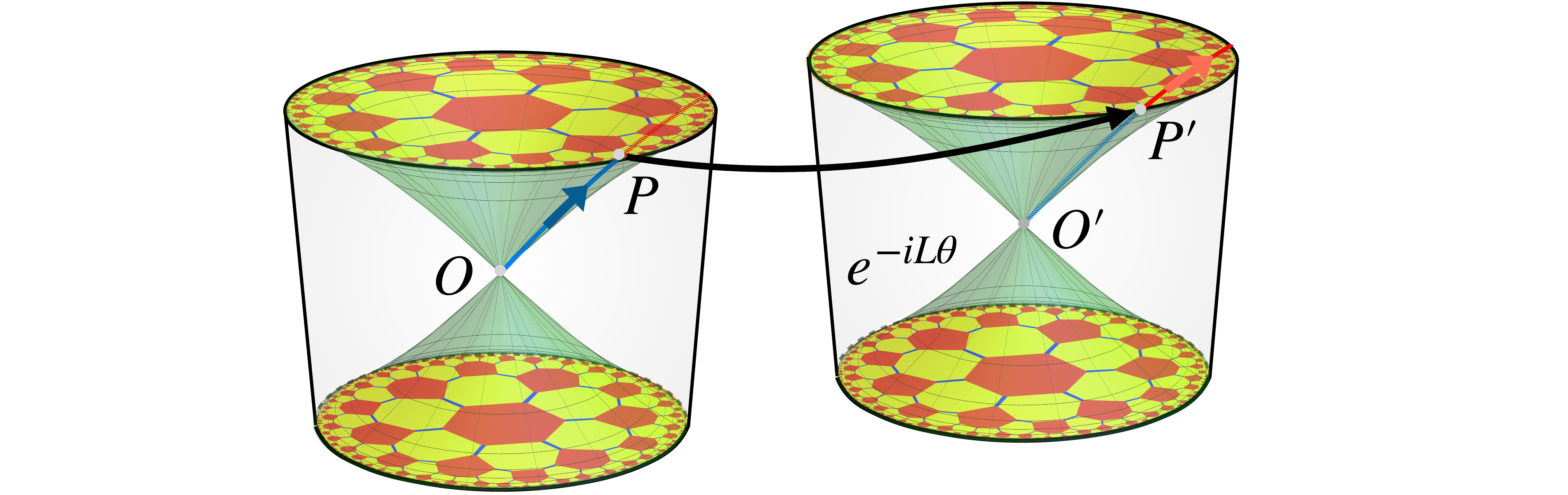}
    \caption
    {
    A schematic  of tunnelings between AdS$_{2+1}$ spacetimes.
    }
    \label{fig:fig3}
\end{figure}
%-------------------------------------------------------------------------

Whereas the change of $\varphi$ is measurable, for instance, by coupling the system to an external quantum spin-valve~\cite{Reiserer2013,Cetina2016,Qi2021}, it is useful to consider platforms in which states with different $\varphi$ cannot be distinguished in experiments. States with the same $\eta$ and $\varphi-\psi$ but different $\varphi$ correspond to a spiral in AdS$_{2+1}$, as shown in Fig.~\ref{fig:fig2}b. 
When experimentalists could only access $\eta$ and $\varphi-\psi$, 
it is desirable to seek the shortest time to transport an initial state $P$ to this spiral. 
Since the spiral and the light cone of $P$ have two intersection points $R_1$ and $R_2$, based on the previous discussions, it takes least time to reach $R_1$ and $R_2$ by using quench dynamics represented by the arrow $PR_1$ or $PR_2$ that lies on the light cone of $P$. 
If one chooses a different quench dynamics such as $PR'$ that corresponds to a constant $\varphi$, it takes a longer time to reach the spiral and access the same observables $\eta$ and $\varphi-\psi$. 
Whereas we focus on the generic case with arbitrary initial and final states, it would be interesting to implement our scheme to dynamical phase transitions, where the initial and final states belong to different quantum phases~\cite{Heyl2013}, for instance, one in the dynamically stable regime and the other in the unstable regime. 

Similar to the conservation of the angular momentum in $SU$(2), the eigenvalue of the Casimir operator, $C=K_0^2-K_1^2-K_2^2$, in $SU$(1,1) is also conserved. Each AdS$_{2+1}$ emergent from quantum dynamics is characterized by a unique $C$. 
If we view an AdS$_{2+1}$ as a universe, trajectories corresponding to quantum dynamics induced by $K_{0,1,2}$ are always confined in the same universe. 
This can be demonstrated using the two-mode squeezing. 
Any state in a single AdS$_{2+1}$ can be written as 
\begin{equation}
\begin{split}
    \ket{\eta,\varphi,\psi;n_0}=&e^{-i(n_0+1)\varphi}\cosh^{-(n_0+1)}(\eta)\sum_{n=0}^\infty \sqrt{\begin{pmatrix}n+n_0\\n\end{pmatrix}}\\
    &\left(-i\tanh(\eta)e^{-i(\varphi-\psi)}\right)^n\ket{n,n+n_0},\label{eq:coh}
\end{split}
\end{equation}
where $n_0$, the difference in the occupation numbers of these two modes, is  a quantum number associated with $C$, and relates to the Bargmann index by $n_0=2k-1$.  
As particles are always created in pairs, one in mode $a$ and the other in mode $b$, $k$ is conserved in quantum dynamics induced by any  $H(\tau)=\sum_{i=0,1,2}\xi_i(\tau)K_i$ and the trajectory is confined within a single AdS$_{2+1}$. 

Once we add an addition operator $L$ outside the $su$(1,1) algebra to the Hamiltonian, $SU$(1,1) symmetry is broken and $k$ is no longer conserved. 
This enlarges the Hilbert space that is relevant in quantum dynamics. 
The quantum system is allowed to move along a trajectory passing through different AdS$_{2+1}$.
For instance, we consider $L=(a^\dag b-b^\dag a)/(2i)$ and the propagators parameterized by $U=e^{-iK_0(\varphi-\psi)}e^{-2iK_1\eta}e^{-iK_0(\varphi+\psi)}e^{-i L\theta}$. 
Let the field coupled to $L$ be $B(\tau)$, such that $Y_3=B(\tau)d\tau/\hbar$, we choose a simple $F=\sqrt{-Y_0^2+Y_1^2+Y_2^2+4Y_3^2}/2$, and the metric is written as 
\begin{equation}
  ds^2=\di\eta^2-\cosh^2(\eta)\di\varphi^2+\sinh^2(\eta)\di\psi^2+\cosh^2(2\eta)\di\theta^2.\label{eq:adses}
\end{equation}
Adding $L$ thus has created a higher dimensional spacetime. 
Whereas a fixed $\theta$ corresponds to a single AdS$_{2+1}$, moving along the $\theta$ direction amounts to tunnelings between different AdS$_{2+1}$ spacetimes. 
In quantum dynamics, such tunnelings provide a geometric description of the couplings between quantum states with different $C$ and Bargmann index, for instance, different $n_0$ in two-mode squeezing. 

In this manuscript, we have used the $SU(1,1)$ group to demonstrate the methodology. 
Our protocol can be directly applied to any other groups. 
The Lie algebra underlying the symmetry group allows us to immediately work out the geometry~\cite{Nielsen2005,Nielsen2006,Jefferson2017,Chapman2018,Guo2018,Susskind2019,Sood2022}. 
The metric tensor then allows experimentalists to access a target state within the least time following the geodesics of the geometry. 
Whereas the spacetime of our universe is determined by the Einstein equation, we have seen that spacetimes can also emerge from quantum dynamics. 
Since such spacetimes are synthetic ones, where the Einstein equation is irrelevant, experimentalists could bypass constraints on the energy-moment tensor and engineer exotic spacetimes, which may be difficult to access in gravity, by manipulating quantum dynamics.
We hope that our work will stimulate more interest in deep connections between quantum dynamics and geometry. 

We acknowledge useful discussions with Chen-Lung Hung, Sergei Khlebnikov, Martin Kruczenski and Nima Lashkari. 
Q.Z. and C.L. are supported by the Air Force Office of Scientific Research under award number FA9550-20-1-0221, DOE QuantISED program of the theory consortium “Intersections of QIS and Theoretical Particle Physics" at Fermilab, and W. M. Keck Foundation.

\onecolumngrid
\newpage
\vspace{0.4in}
\centerline{\bf\large Supplementary Materials for }
\centerline{\bf\large ``Emergent spacetimes 
from Hermitian and non-Hermitian quantum dynamics"}
\setcounter{equation}{0}
\setcounter{figure}{0}
\setcounter{table}{0}
\makeatletter
\renewcommand{\theequation}{S\arabic{equation}}
\renewcommand{\thefigure}{S\arabic{figure}}
\renewcommand{\thetable}{S\arabic{table}}
\vspace{0.2in}

\subsection{Squashed or stretched AdS, AdS black hole and soliton}
In the main text, we have been focusing on an isotropic and homogeneous definition of proper length $F$, which corresponds to a constant field strength $\xi=\sqrt{\xi_0^2-\xi_1^2-\xi_2^2}$. 
In certain systems, it might be easier or more difficult to implement some gates than others. 
It is thus desirable to consider anisotropic definition of proper length, which is also known as anisotropic cost functions~\cite{SSusskind2019}. 
To this end, we first consider a generalized form of the anisotropic cost function 
$\tilde{F}=\sqrt{-\lambda Y_0^2+Y_1^2+Y_2^2}/2$, where $\lambda\neq1$. 
Similar to the $SU$(2) case where such deformation leads to a Berger sphere~\cite{SSusskind2019,SBengtsson2006}, we obtain a squashed or stretched AdS$_{2+1}$ spacetime along timelike directions,
\begin{equation}
    ds^2=d\eta^2+\sinh^2(\eta)d\psi^2-\cosh^2(\eta)d\varphi^2+(1-\lambda)(\sinh^2(\eta)d\psi+\cosh^2(\eta)d\varphi)^2.
    \label{eq:BergerAdS}
\end{equation}
In addition, it is of interest to explore a non-constant field strength that often leads to exotic physics. 
For instance, in the $SU$(2) case, a magnetic field that varies over a sphere gives rise to 't Hooft-Polyakov monopole~\cite{StHooft1974,SPolyakov1974}. 
Here, a non-constant field and the corresponding inhomogeneous cost function produces intriguing spacetime metrics. 
We emphasize that, in addition to theoretical interest, it is an important question to consider inhomogeneous cost function, as it may become easier or harder to apply certain gates when the system evolves to a different state~\cite{SSusskind2019}. 
For instance, in a driven BEC, all three components, $\xi_{i=0,1,2}$, depend on the condensate density. 
Whereas the condensate density can be well approximated by a constant at small times, at large times, the occupation at excited states becomes significant such that the condensate density, as well as the field strength $\xi$, cannot be treated as a constant anymore. 

As an example, we  consider an inhomogeneous cost function, $\tilde{F}=f^{1/2}(\eta,\psi,\varphi)\sqrt{-Y_0^2+Y_1^2+Y_2^2}/2$, which corresponds to an isotropic but non-uniform field strength in the parameter space. 
Using the same method as presented in the main text, we obtain the following metric 
\begin{equation}
  ds^2=f(\eta,\psi,\varphi)\left(\di\eta^2-\cosh(\eta)^2\di\varphi^2+\sinh(\eta)^2\di\psi^2\right).\label{eq:adsc}
\end{equation}
Manipulating $f(\eta,\psi,\varphi)$ provides us with different spacetimes.
For instance, when $f(\eta,\psi,\varphi)=({\rho}/z_H)^2\sinh^{-2}({\rho}/z_H)$, where ${\rho}(\eta,\psi,\varphi)=(\cosh(\eta)\cos(\varphi)-\sinh(\eta)\sin(\psi))^{-1}$, the spacetime has an event horizon located at $z_H$. 
This can be seen from a coordinate transformation into the horospheric coordinates, defined as 
\begin{equation}
    \rho=\frac1{\cosh(\eta)\cos(\varphi)-\sinh(\eta)\sin(\psi)},\quad 
    x=\frac{\sinh(\eta)\cos(\psi)}{\cosh(\eta)\cos(\varphi)-\sinh(\eta)\sin(\psi)},\quad 
    t=\frac{\cosh(\eta)\sin(\varphi)}{\cosh(\eta)\cos(\varphi)-\sinh(\eta)\sin(\psi)},
\end{equation}
and the resulted metric reads 
\begin{equation}
  ds^2=(1/z_H)^2\sinh^{-2}(\rho/z_H)\left(-\di t^2+\di x^2+\di \rho^2\right).\label{eqs:adsc2}
\end{equation}
We further consider a coordinate transformation of $\rho$, $z/z_H=\tanh(\rho/z_H)$, such that
\begin{equation}
  ds^2=-\left(\frac1{z^2}-\frac1{z_H^2}\right)\di t^2+\frac1{z^2\left(1-z^2/z_H^2\right)}\di z^2+\left(\frac1{z^2}-\frac1{z_H^2}\right)\di x^2,\label{M1}
\end{equation}
$\rho\in(0,\infty)$ is mapped to $z\in(0,z_H)$.
When $z=z_H$, the speed of light vanishes as a characteristic feature of event horizon. 
This metric is related to that of BTZ black hole~\cite{SAuzzi2020}, 
\begin{equation}
   ds^2=-\left(\frac1{z^2}-\frac1{z_H^2}\right)\di t^2+\frac1{z^2(1-z^2/z_H^2)}\di z^2+\frac{1}{z^2}\di x^2.\label{M2}
\end{equation}
It is apparent that Eq.(\ref{M1}) and Eq.(\ref{M2}) become identical when $dx=0$, i.e., for any fixed $x$.   
Eq.(\ref{M1}) is also related to the  AdS$_{2+1}$ soliton metric~\cite{SReynolds2018},
\begin{equation}
    ds^2=-\frac{1}{z^2}\di t^2+\frac1{z^2(1-z^2/z_H^2)}\di z^2+\left(\frac1{z^2}-\frac1{z_H^2}\right)\di x^2.\label{M3}
\end{equation}
Eq.(\ref{M3}) and Eq.(\ref{M1}) coincide when $dt=0$. 

\subsection{Geodesic equations in  AdS$_{2+1}$ spacetime}
The geodesic equation for a metric tensor ${\bf g}=g_{\mu\nu}\di x^{\mu}\di x^{\nu}$ is written as~\cite{SCarroll2019}
\begin{equation}
    \frac{d^2x^{\mu}}{ds^2}=-\Gamma^{\mu}_{\nu\lambda}\frac{dx^{\nu}}{ds}\frac{dx^{\lambda}}{ds}.
\end{equation}
where $\Gamma^{\mu}_{\nu\lambda}$ are Christoffel symbols. 
For the metric of  AdS$_{2+1}$ spacetime in coordinate shown in Eq.(6) of the main text, non-vanishing Christoffel symbols read
\begin{equation}
  \Gamma^{\varphi}_{\varphi\eta}=\tanh(\eta),\quad \Gamma^{\eta}_{\varphi\varphi}=\sinh(\eta)\cosh(\eta),\quad 
  \Gamma^{\eta}_{\psi\psi}=-\sinh(\eta)\cosh(\eta),\quad \Gamma^{\psi}_{\eta\psi}=\coth(\eta).
\end{equation}
The geodesic equations are written as  
\begin{equation}
  \begin{split}
    &\frac{d^2\varphi}{ds^2}=-2\tanh(\eta)\frac{d\varphi}{ds}\frac{d\eta}{ds},\\
    &\frac{d^2\eta}{ds^2}=-\sinh(\eta)\cosh(\eta)\left[\left(\frac{d\varphi}{ds}\right)^2-\left(\frac{d\psi}{ds}\right)^2\right],\\
    &\frac{d^2\psi}{ds^2}=-2\coth(\eta)\frac{d\psi}{ds}\frac{d\eta}{ds}.
  \end{split}\label{S-geo}
\end{equation}
Since $\xi$ is a constant for quench dynamics, we can replace $s$ by $\tau$ in Eq.~(\ref{S-geo}). 
Therefore, Eq.~(9) of the main text is recovered.

\subsection{Deviations from the shortest path}
We consider a path $OP'R$ where $P'$ is no longer on the light cone. 
When $P'$ is in the timelike regime, the Hamiltonian is written as 
\begin{equation}
    \begin{split}
        H_1 =& \big[(\delta+\delta^{-1})K_0-(\delta-\delta^{-1})K_1\big]\hbar\omega_0,\qquad\qquad\qquad\,\,\, 0<\tau<\tau_1,\\
        H_2 =& \frac{i\hbar}{\tau_2}\log(e^{-i(K_0-K_1)\omega_0T_2}e^{-i(K_0+K_1)\omega_0T_1}e^{iH_1\tau_1/\hbar}),\quad \tau_1<\tau<\tau_2.\\
    \end{split}
\end{equation}
For any given $\delta$ and $\tau_1$, an appropriate $H_2$ defined in the above equation can be chosen such that the same target state at $R$ can be accessed at a certain time $\tau_2$. 
When $\delta$ decreases down to zero, the field strength remains constant. 
Choosing an appropriate $\tau_1$, $P'$ approaches $P$ and we recover the shortest path in the limit where $\delta\rightarrow 0$. 

To compute the proper length $S$, or equivalently, the time spent to reach $R$ from $O$, as a function of $\delta$ and $\tau_1$, the simplest means is to implement the representation using spin-1/2 in a complex magnetic field, as shown in the last row of Table~I in the main text. 
For instance, the propagator along the shortest path can be written as $U_{OPR}=e^{-i\sigma_-\omega_0T_2}e^{-i\sigma_+\omega_0T_1}$, corresponding to dynamics at the exceptional points in non-Hermitian physics and the trajectory on the light cones in AdS$_{2+1}$, where $\sigma_\pm=(\sigma_x\pm i\sigma_y)/2$. 
For generic $\delta$ and $\tau_1$, the analytical expression for the proper length reads 
\begin{equation}
    S(\delta,\tau_1)/i =\omega_0 \tau_1+
   \bigg|\frac1{2i}\log(\frac{f(\delta,\tau_1)-\sqrt{f(\delta,\tau_1)^2-\delta^2}}{f(\delta,\tau_1) + \sqrt{f(\delta,\tau_1)^2-\delta^2}})\bigg|,\label{Scost}
\end{equation}
where $f(\delta,\tau_1)=[(2 - \omega_0^2T_1T_2) \delta \cos(\omega_0\tau_1) + \omega_0(T_2 + \delta^2T_1) \sin(\omega_0\tau_1)]/2$. 
When $\delta\neq 0$, $S$ is finite.
When $\delta$ decreases down to zero and an appropriate $\tau_1= T_1 \delta$ is chosen, $S$ approaches zero, as we can see from Eq.~(\ref{Scost}) $\lim_{\delta\to 0}S(\delta, T_1\delta)/i=\lim_{\delta\to 0}(\omega_0T_1+\sqrt{\omega_0^4T_1^3T_2/3})\delta+O(\delta^3)=0$. 
This is expected as the trajectory approaches the shortest path $OPR$.

We also consider that $P'$ falls in the spacelike regime. The Hamiltonian is written as 
\begin{equation}
    \begin{split}
        H_1 =& \big[-(\delta-\delta^{-1})K_0+(\delta+\delta^{-1})K_1\big]\hbar\omega_0,\qquad\qquad\qquad\,\,\, 0<\tau<\tau_1,\\
        H_2 =& \frac{i\hbar}{\tau_2}\log(e^{-i(K_0-K_1)\omega_0T_2}e^{-i(K_0+K_1)\omega_0T_1}e^{iH_1\tau_1/\hbar}),\qquad \tau_1<\tau<\tau_2.\\
    \end{split}
\end{equation}
Its proper length can be obtained by analytically continuing $\delta$ and $\tau_1$ in Eq.~\ref{Scost} to the complex domain,
\begin{equation}
    S(i\delta,i\tau_1)= \omega_0 \tau_1+
  \bigg| \frac1{2} \log(\frac{f(i\delta,i\tau_1)-\sqrt{f(i\delta,i\tau_1)^2+\delta^2}}{f(i\delta,i\tau_1) + \sqrt{f(i\delta,i\tau_1)^2+\delta^2}})\bigg|.\label{Scost2}
\end{equation}


\begin{thebibliography}{50}%
\makeatletter
\providecommand \@ifxundefined [1]{%
 \@ifx{#1\undefined}
}%
\providecommand \@ifnum [1]{%
 \ifnum #1\expandafter \@firstoftwo
 \else \expandafter \@secondoftwo
 \fi
}%
\providecommand \@ifx [1]{%
 \ifx #1\expandafter \@firstoftwo
 \else \expandafter \@secondoftwo
 \fi
}%
\providecommand \natexlab [1]{#1}%
\providecommand \enquote  [1]{#1}%
\providecommand \bibnamefont  [1]{#1}%
\providecommand \bibfnamefont [1]{#1}%
\providecommand \citenamefont [1]{#1}%
\providecommand \href@noop [0]{\@secondoftwo}%
\providecommand \href [0]{\begingroup \@sanitize@url \@href}%
\providecommand \@href[1]{\@@startlink{#1}\@@href}%
\providecommand \@@href[1]{\endgroup#1\@@endlink}%
\providecommand \@sanitize@url [0]{\catcode `\\12\catcode `\$12\catcode
  `\&12\catcode `\#12\catcode `\^12\catcode `\_12\catcode `\%12\relax}%
\providecommand \@@startlink[1]{}%
\providecommand \@@endlink[0]{}%
\providecommand \url  [0]{\begingroup\@sanitize@url \@url }%
\providecommand \@url [1]{\endgroup\@href {#1}{\urlprefix }}%
\providecommand \urlprefix  [0]{URL }%
\providecommand \Eprint [0]{\href }%
\providecommand \doibase [0]{https://dx.doi.org}%
\providecommand \selectlanguage [0]{\@gobble}%
\providecommand \bibinfo  [0]{\@secondoftwo}%
\providecommand \bibfield  [0]{\@secondoftwo}%
\providecommand \translation [1]{[#1]}%
\providecommand \BibitemOpen [0]{}%
\providecommand \bibitemStop [0]{}%
\providecommand \bibitemNoStop [0]{.\EOS\space}%
\providecommand \EOS [0]{\spacefactor3000\relax}%
\providecommand \BibitemShut  [1]{\csname bibitem#1\endcsname}%
\let\auto@bib@innerbib\@empty
%</preamble>
\bibitem [{\citenamefont {Maldacena}(1998)}]{Maldacena1998}%
  \BibitemOpen
  \bibfield  {author} {\bibinfo {author} {\bibfnamefont {J.}~\bibnamefont
  {Maldacena}},\ }\bibfield  {title} {\bibinfo {title} {The large \textit{N}
  limit of superconformal field theories and supergravity},\ }\href
  {\doibase/10.4310/atmp.1998.v2.n2.a1} {\bibfield  {journal} {\bibinfo
  {journal} {Advances in Theoretical and Mathematical Physics}\ }\textbf
  {\bibinfo {volume} {2}},\ \bibinfo {pages} {231} (\bibinfo {year}
  {1998})}\BibitemShut {NoStop}%
\bibitem [{\citenamefont {Gubser}\ \emph {et~al.}(1998)\citenamefont {Gubser},
  \citenamefont {Klebanov},\ and\ \citenamefont {Polyakov}}]{Gubser1998}%
  \BibitemOpen
  \bibfield  {author} {\bibinfo {author} {\bibfnamefont {S.}~\bibnamefont
  {Gubser}}, \bibinfo {author} {\bibfnamefont {I.}~\bibnamefont {Klebanov}}, \
  and\ \bibinfo {author} {\bibfnamefont {A.}~\bibnamefont {Polyakov}},\
  }\bibfield  {title} {\bibinfo {title} {Gauge theory correlators from
  non-critical string theory},\ }\href {\doibase/10.1016/s0370-2693(98)00377-3}
  {\bibfield  {journal} {\bibinfo  {journal} {Physics Letters B}\ }\textbf
  {\bibinfo {volume} {428}},\ \bibinfo {pages} {105} (\bibinfo {year}
  {1998})}\BibitemShut {NoStop}%
\bibitem [{\citenamefont {Witten}(1998)}]{Witten1998}%
  \BibitemOpen
  \bibfield  {author} {\bibinfo {author} {\bibfnamefont {E.}~\bibnamefont
  {Witten}},\ }\bibfield  {title} {\bibinfo {title} {Anti de Sitter space and
  holography},\ }\href {\doibase/10.4310/atmp.1998.v2.n2.a2} {\bibfield
  {journal} {\bibinfo  {journal} {Advances in Theoretical and Mathematical
  Physics}\ }\textbf {\bibinfo {volume} {2}},\ \bibinfo {pages} {253} (\bibinfo
  {year} {1998})}\BibitemShut {NoStop}%
\bibitem [{\citenamefont {Vidal}(2007)}]{Vidal2007}%
  \BibitemOpen
  \bibfield  {author} {\bibinfo {author} {\bibfnamefont {G.}~\bibnamefont
  {Vidal}},\ }\bibfield  {title} {\bibinfo {title} {Entanglement
  renormalization},\ }\href {\doibase/10.1103/PhysRevLett.99.220405} {\bibfield
   {journal} {\bibinfo  {journal} {Phys. Rev. Lett.}\ }\textbf {\bibinfo
  {volume} {99}},\ \bibinfo {pages} {220405} (\bibinfo {year}
  {2007})}\BibitemShut {NoStop}%
\bibitem [{\citenamefont {Swingle}(2012)}]{Brian2012}%
  \BibitemOpen
  \bibfield  {author} {\bibinfo {author} {\bibfnamefont {B.}~\bibnamefont
  {Swingle}},\ }\bibfield  {title} {\bibinfo {title} {Entanglement
  renormalization and holography},\ }\href
  {\doibase/10.1103/PhysRevD.86.065007} {\bibfield  {journal} {\bibinfo
  {journal} {Phys. Rev. D}\ }\textbf {\bibinfo {volume} {86}},\ \bibinfo
  {pages} {065007} (\bibinfo {year} {2012})}\BibitemShut {NoStop}%
\bibitem [{\citenamefont {Nozaki}\ \emph {et~al.}(2012)\citenamefont {Nozaki},
  \citenamefont {Ryu},\ and\ \citenamefont {Takayanagi}}]{Nozaki2012}%
  \BibitemOpen
  \bibfield  {author} {\bibinfo {author} {\bibfnamefont {M.}~\bibnamefont
  {Nozaki}}, \bibinfo {author} {\bibfnamefont {S.}~\bibnamefont {Ryu}}, \ and\
  \bibinfo {author} {\bibfnamefont {T.}~\bibnamefont {Takayanagi}},\ }\bibfield
   {title} {\bibinfo {title} {Holographic geometry of entanglement
  renormalization in quantum field theories},\ }\href
  {\doibase/10.1007/jhep10(2012)193} {\bibfield  {journal} {\bibinfo  {journal}
  {Journal of High Energy Physics}\ }\textbf {\bibinfo {volume} {2012}}
  (\bibinfo {year} {2012}),\ 10.1007/jhep10(2012)193}\BibitemShut {NoStop}%
\bibitem [{\citenamefont {Qi}(2013)}]{Qi2013}%
  \BibitemOpen
  \bibfield  {author} {\bibinfo {author} {\bibfnamefont {X.-L.}\ \bibnamefont
  {Qi}},\ }\href {\doibase/10.48550/ARXIV.1309.6282} {\bibinfo {title} {Exact
  holographic mapping and emergent space-time geometry},\ arXiv:1309.6282} (\bibinfo {year}
  {2013})\BibitemShut {NoStop}%
\bibitem [{\citenamefont {Pastawski}\ \emph {et~al.}(2015)\citenamefont
  {Pastawski}, \citenamefont {Yoshida}, \citenamefont {Harlow},\ and\
  \citenamefont {Preskill}}]{Preskill2015}%
  \BibitemOpen
  \bibfield  {author} {\bibinfo {author} {\bibfnamefont {F.}~\bibnamefont
  {Pastawski}}, \bibinfo {author} {\bibfnamefont {B.}~\bibnamefont {Yoshida}},
  \bibinfo {author} {\bibfnamefont {D.}~\bibnamefont {Harlow}}, \ and\ \bibinfo
  {author} {\bibfnamefont {J.}~\bibnamefont {Preskill}},\ }\bibfield  {title}
  {\bibinfo {title} {Holographic quantum error-correcting codes: toy models for
  the bulk/boundary correspondence},\ }\href {\doibase/10.1007/jhep06(2015)149}
  {\bibfield  {journal} {\bibinfo  {journal} {Journal of High Energy Physics}\
  }\textbf {\bibinfo {volume} {2015}} (\bibinfo {year} {2015}),\
  10.1007/jhep06(2015)149}\BibitemShut {NoStop}%
\bibitem [{\citenamefont {Nielsen}(2005)}]{Nielsen2005}%
  \BibitemOpen
  \bibfield  {author} {\bibinfo {author} {\bibfnamefont {M.~A.}\ \bibnamefont
  {Nielsen}},\ }\href {\doibase/10.48550/ARXIV.QUANT-PH/0502070} {\bibinfo
  {title} {A geometric approach to quantum circuit lower bounds},\ arXiv:quant-ph/0502070} (\bibinfo
  {year} {2005})\BibitemShut {NoStop}%
\bibitem [{\citenamefont {Nielsen}\ \emph {et~al.}(2006)\citenamefont
  {Nielsen}, \citenamefont {Dowling}, \citenamefont {Gu},\ and\ \citenamefont
  {Doherty}}]{Nielsen2006}%
  \BibitemOpen
  \bibfield  {author} {\bibinfo {author} {\bibfnamefont {M.~A.}\ \bibnamefont
  {Nielsen}}, \bibinfo {author} {\bibfnamefont {M.~R.}\ \bibnamefont
  {Dowling}}, \bibinfo {author} {\bibfnamefont {M.}~\bibnamefont {Gu}}, \ and\
  \bibinfo {author} {\bibfnamefont {A.~C.}\ \bibnamefont {Doherty}},\
  }\bibfield  {title} {\bibinfo {title} {Quantum computation as geometry},\
  }\href {\doibase/10.1126/science.1121541} {\bibfield  {journal} {\bibinfo
  {journal} {Science}\ }\textbf {\bibinfo {volume} {311}},\ \bibinfo {pages}
  {1133} (\bibinfo {year} {2006})}\BibitemShut {NoStop}%
\bibitem [{\citenamefont {Jefferson}\ and\ \citenamefont
  {Myers}(2017)}]{Jefferson2017}%
  \BibitemOpen
  \bibfield  {author} {\bibinfo {author} {\bibfnamefont {R.~A.}\ \bibnamefont
  {Jefferson}}\ and\ \bibinfo {author} {\bibfnamefont {R.~C.}\ \bibnamefont
  {Myers}},\ }\bibfield  {title} {\bibinfo {title} {Circuit complexity in
  quantum field theory},\ }\href {\doibase/10.1007/jhep10(2017)107} {\bibfield
  {journal} {\bibinfo  {journal} {Journal of High Energy Physics}\ }\textbf
  {\bibinfo {volume} {2017}} (\bibinfo {year} {2017}),\
  10.1007/jhep10(2017)107}\BibitemShut {NoStop}%
\bibitem [{\citenamefont {Chapman}\ \emph {et~al.}(2018)\citenamefont
  {Chapman}, \citenamefont {Heller}, \citenamefont {Marrochio},\ and\
  \citenamefont {Pastawski}}]{Chapman2018}%
  \BibitemOpen
  \bibfield  {author} {\bibinfo {author} {\bibfnamefont {S.}~\bibnamefont
  {Chapman}}, \bibinfo {author} {\bibfnamefont {M.~P.}\ \bibnamefont {Heller}},
  \bibinfo {author} {\bibfnamefont {H.}~\bibnamefont {Marrochio}}, \ and\
  \bibinfo {author} {\bibfnamefont {F.}~\bibnamefont {Pastawski}},\ }\bibfield
  {title} {\bibinfo {title} {Toward a definition of complexity for quantum
  field theory states},\ }\href {\doibase/10.1103/PhysRevLett.120.121602}
  {\bibfield  {journal} {\bibinfo  {journal} {Phys. Rev. Lett.}\ }\textbf
  {\bibinfo {volume} {120}},\ \bibinfo {pages} {121602} (\bibinfo {year}
  {2018})}\BibitemShut {NoStop}%
\bibitem [{\citenamefont {Guo}\ \emph {et~al.}(2018)\citenamefont {Guo},
  \citenamefont {Hernandez}, \citenamefont {Myers},\ and\ \citenamefont
  {Ruan}}]{Guo2018}%
  \BibitemOpen
  \bibfield  {author} {\bibinfo {author} {\bibfnamefont {M.}~\bibnamefont
  {Guo}}, \bibinfo {author} {\bibfnamefont {J.}~\bibnamefont {Hernandez}},
  \bibinfo {author} {\bibfnamefont {R.~C.}\ \bibnamefont {Myers}}, \ and\
  \bibinfo {author} {\bibfnamefont {S.-M.}\ \bibnamefont {Ruan}},\ }\bibfield
  {title} {\bibinfo {title} {Circuit complexity for coherent states},\ }\href
  {\doibase/10.1007/jhep10(2018)011} {\bibfield  {journal} {\bibinfo  {journal}
  {Journal of High Energy Physics}\ }\textbf {\bibinfo {volume} {2018}}
  (\bibinfo {year} {2018}),\ 10.1007/jhep10(2018)011}\BibitemShut {NoStop}%
\bibitem [{\citenamefont {Brown}\ and\ \citenamefont
  {Susskind}(2019)}]{Susskind2019}%
  \BibitemOpen
  \bibfield  {author} {\bibinfo {author} {\bibfnamefont {A.~R.}\ \bibnamefont
  {Brown}}\ and\ \bibinfo {author} {\bibfnamefont {L.}~\bibnamefont
  {Susskind}},\ }\bibfield  {title} {\bibinfo {title} {Complexity geometry of a
  single qubit},\ }\href {\doibase/10.1103/PhysRevD.100.046020} {\bibfield
  {journal} {\bibinfo  {journal} {Phys. Rev. D}\ }\textbf {\bibinfo {volume}
  {100}},\ \bibinfo {pages} {046020} (\bibinfo {year} {2019})}\BibitemShut
  {NoStop}%
\bibitem [{\citenamefont {Sood}\ and\ \citenamefont
  {Kruczenski}(2022)}]{Sood2022}%
  \BibitemOpen
  \bibfield  {author} {\bibinfo {author} {\bibfnamefont {U.}~\bibnamefont
  {Sood}}\ and\ \bibinfo {author} {\bibfnamefont {M.}~\bibnamefont
  {Kruczenski}},\ }\bibfield  {title} {\bibinfo {title} {Circuit complexity
  near critical points},\ }\href {\doibase/10.1088/1751-8121/ac5b8f} {\bibfield
   {journal} {\bibinfo  {journal} {Journal of Physics A: Mathematical and
  Theoretical}\ }\textbf {\bibinfo {volume} {55}},\ \bibinfo {pages} {185301}
  (\bibinfo {year} {2022})}\BibitemShut {NoStop}%
\bibitem [{\citenamefont {Lv}\ \emph {et~al.}(2020)\citenamefont {Lv},
  \citenamefont {Zhang},\ and\ \citenamefont {Zhou}}]{echo1}%
  \BibitemOpen
  \bibfield  {author} {\bibinfo {author} {\bibfnamefont {C.}~\bibnamefont
  {Lv}}, \bibinfo {author} {\bibfnamefont {R.}~\bibnamefont {Zhang}}, \ and\
  \bibinfo {author} {\bibfnamefont {Q.}~\bibnamefont {Zhou}},\ }\bibfield
  {title} {\bibinfo {title} {$su(1,1)$ echoes for breathers in quantum gases},\
  }\href {\doibase/10.1103/PhysRevLett.125.253002} {\bibfield  {journal}
  {\bibinfo  {journal} {Phys. Rev. Lett.}\ }\textbf {\bibinfo {volume} {125}},\
  \bibinfo {pages} {253002} (\bibinfo {year} {2020})}\BibitemShut {NoStop}%
\bibitem [{\citenamefont {Lyu}\ \emph {et~al.}(2020)\citenamefont {Lyu},
  \citenamefont {Lv},\ and\ \citenamefont {Zhou}}]{echo2}%
  \BibitemOpen
  \bibfield  {author} {\bibinfo {author} {\bibfnamefont {C.}~\bibnamefont
  {Lyu}}, \bibinfo {author} {\bibfnamefont {C.}~\bibnamefont {Lv}}, \ and\
  \bibinfo {author} {\bibfnamefont {Q.}~\bibnamefont {Zhou}},\ }\bibfield
  {title} {\bibinfo {title} {Geometrizing quantum dynamics of a Bose-Einstein
  condensate},\ }\href {\doibase/10.1103/PhysRevLett.125.253401} {\bibfield
  {journal} {\bibinfo  {journal} {Phys. Rev. Lett.}\ }\textbf {\bibinfo
  {volume} {125}},\ \bibinfo {pages} {253401} (\bibinfo {year}
  {2020})}\BibitemShut {NoStop}%
\bibitem [{\citenamefont {Chen}\ \emph {et~al.}(2020)\citenamefont {Chen},
  \citenamefont {Zhang}, \citenamefont {Zheng}, \citenamefont {Wu},\ and\
  \citenamefont {Zhai}}]{echo3}%
  \BibitemOpen
  \bibfield  {author} {\bibinfo {author} {\bibfnamefont {Y.-Y.}\ \bibnamefont
  {Chen}}, \bibinfo {author} {\bibfnamefont {P.}~\bibnamefont {Zhang}},
  \bibinfo {author} {\bibfnamefont {W.}~\bibnamefont {Zheng}}, \bibinfo
  {author} {\bibfnamefont {Z.}~\bibnamefont {Wu}}, \ and\ \bibinfo {author}
  {\bibfnamefont {H.}~\bibnamefont {Zhai}},\ }\bibfield  {title} {\bibinfo
  {title} {Many-body echo},\ }\href {\doibase/10.1103/PhysRevA.102.011301}
  {\bibfield  {journal} {\bibinfo  {journal} {Phys. Rev. A}\ }\textbf {\bibinfo
  {volume} {102}},\ \bibinfo {pages} {011301(R)} (\bibinfo {year}
  {2020})}\BibitemShut {NoStop}%
\bibitem [{\citenamefont {Cheng}\ and\ \citenamefont {Shi}(2021)}]{Cheng2021}%
  \BibitemOpen
  \bibfield  {author} {\bibinfo {author} {\bibfnamefont {Y.}~\bibnamefont
  {Cheng}}\ and\ \bibinfo {author} {\bibfnamefont {Z.-Y.}\ \bibnamefont
  {Shi}},\ }\bibfield  {title} {\bibinfo {title} {Many-body dynamics with
  time-dependent interaction},\ }\href {\doibase/10.1103/PhysRevA.104.023307}
  {\bibfield  {journal} {\bibinfo  {journal} {Phys. Rev. A}\ }\textbf {\bibinfo
  {volume} {104}},\ \bibinfo {pages} {023307} (\bibinfo {year}
  {2021})}\BibitemShut {NoStop}%
\bibitem [{\citenamefont {Zhang}\ \emph {et~al.}(2022)\citenamefont {Zhang},
  \citenamefont {Yang}, \citenamefont {Lv}, \citenamefont {Ma},\ and\
  \citenamefont {Zhang}}]{Zhang2022}%
  \BibitemOpen
  \bibfield  {author} {\bibinfo {author} {\bibfnamefont {J.}~\bibnamefont
  {Zhang}}, \bibinfo {author} {\bibfnamefont {X.}~\bibnamefont {Yang}},
  \bibinfo {author} {\bibfnamefont {C.}~\bibnamefont {Lv}}, \bibinfo {author}
  {\bibfnamefont {S.}~\bibnamefont {Ma}}, \ and\ \bibinfo {author}
  {\bibfnamefont {R.}~\bibnamefont {Zhang}},\ }\href
  {\doibase/10.48550/ARXIV.2201.04304} {\bibinfo {title} {Quantum dynamics of
  cold atomic gas with $su(1, 1)$ symmetry},\ arXiv: 2201.04304} (\bibinfo {year}
  {2022})\BibitemShut {NoStop}%
\bibitem [{\citenamefont {Puri}(2001)}]{Puri2001}%
  \BibitemOpen
  \bibfield  {author} {\bibinfo {author} {\bibfnamefont {R.~R.}\ \bibnamefont
  {Puri}},\ }\href@noop {} { {\bibinfo {title} {Mathematical methods of
  quantum optics}}},\ Vol.~\bibinfo {volume} {79}\ (\bibinfo  {publisher}
  {Springer Science \& Business Media},\ \bibinfo {year} {2001})\BibitemShut
  {NoStop}%
\bibitem [{\citenamefont {Gilmore}(2008)}]{Gilmore2008}%
  \BibitemOpen
  \bibfield  {author} {\bibinfo {author} {\bibfnamefont {R.}~\bibnamefont
  {Gilmore}},\ }\href@noop {} { {\bibinfo {title} {Lie groups, physics,
  and geometry: an introduction for physicists, engineers and chemists}}}\
  (\bibinfo  {publisher} {Cambridge University Press},\ \bibinfo {year}
  {2008})\BibitemShut {NoStop}%
\bibitem [{\citenamefont {Bengtsson}\ and\ \citenamefont
  {Sandin}(2006)}]{Bengtsson2006}%
  \BibitemOpen
  \bibfield  {author} {\bibinfo {author} {\bibfnamefont {I.}~\bibnamefont
  {Bengtsson}}\ and\ \bibinfo {author} {\bibfnamefont {P.}~\bibnamefont
  {Sandin}},\ }\bibfield  {title} {\bibinfo {title} {Anti-de Sitter space,
  squashed and stretched},\ }\href {\doibase/10.1088/0264-9381/23/3/022}
  {\bibfield  {journal} {\bibinfo  {journal} {Classical and Quantum Gravity}\
  }\textbf {\bibinfo {volume} {23}},\ \bibinfo {pages} {971} (\bibinfo {year}
  {2006})}\BibitemShut {NoStop}%
\bibitem [{\citenamefont {Morita}(2019)}]{Morita2019}%
\BibitemOpen
\bibfield  {author} {\bibinfo {author} {\bibfnamefont {T.}~\bibnamefont
{Morita}},\ }\bibfield  {title} {\bibinfo {title} {Thermal emission from
semiclassical dynamical systems},\ }\href
{https://doi.org/10.1103/PhysRevLett.122.101603} {\bibfield  {journal}
{\bibinfo  {journal} {Phys. Rev. Lett.}\ }\textbf {\bibinfo {volume} {122}},\
\bibinfo {pages} {101603} (\bibinfo {year} {2019})}\BibitemShut {NoStop}%
\bibitem [{\citenamefont {Tian}\ \emph {et~al.}(2022)\citenamefont {Tian},
\citenamefont {Lin}, \citenamefont {Fischer},\ and\ \citenamefont
{Du}}]{Tian2022}%
\BibitemOpen
\bibfield  {author} {\bibinfo {author} {\bibfnamefont {Z.}~\bibnamefont
{Tian}}, \bibinfo {author} {\bibfnamefont {Y.}~\bibnamefont {Lin}}, \bibinfo
{author} {\bibfnamefont {U.~R.}\ \bibnamefont {Fischer}},\ and\ \bibinfo
{author} {\bibfnamefont {J.}~\bibnamefont {Du}},\ }\bibfield  {title}
{\bibinfo {title} {Testing the upper bound on the speed of scrambling with an
analogue of Hawking radiation using trapped ions},\ }\bibfield  {journal}
{\bibinfo  {journal} {The European Physical Journal C}\ }\textbf {\bibinfo
{volume} {82}},\ \href {https://doi.org/10.1140/epjc/s10052-022-10149-8}
{10.1140/epjc/s10052-022-10149-8} (\bibinfo {year} {2022})\BibitemShut
{NoStop}%
\bibitem [{\citenamefont {Pitaevskii}\ and\ \citenamefont
  {Rosch}(1997)}]{Pitaevskii1997}%
  \BibitemOpen
  \bibfield  {author} {\bibinfo {author} {\bibfnamefont {L.~P.}\ \bibnamefont
  {Pitaevskii}}\ and\ \bibinfo {author} {\bibfnamefont {A.}~\bibnamefont
  {Rosch}},\ }\bibfield  {title} {\bibinfo {title} {Breathing modes and hidden
  symmetry of trapped atoms in two dimensions},\ }\href
  {\doibase/10.1103/PhysRevA.55.R853} {\bibfield  {journal} {\bibinfo
  {journal} {Phys. Rev. A}\ }\textbf {\bibinfo {volume} {55}},\ \bibinfo
  {pages} {R853} (\bibinfo {year} {1997})}\BibitemShut {NoStop}%
\bibitem [{\citenamefont {Saint-Jalm}\ \emph {et~al.}(2019)\citenamefont
  {Saint-Jalm}, \citenamefont {Castilho}, \citenamefont {Le~Cerf},
  \citenamefont {Bakkali-Hassani}, \citenamefont {Ville}, \citenamefont
  {Nascimbene}, \citenamefont {Beugnon},\ and\ \citenamefont
  {Dalibard}}]{Dalibard2019}%
  \BibitemOpen
  \bibfield  {author} {\bibinfo {author} {\bibfnamefont {R.}~\bibnamefont
  {Saint-Jalm}}, \bibinfo {author} {\bibfnamefont {P.~C.~M.}\ \bibnamefont
  {Castilho}}, \bibinfo {author} {\bibfnamefont {E.}~\bibnamefont {Le~Cerf}},
  \bibinfo {author} {\bibfnamefont {B.}~\bibnamefont {Bakkali-Hassani}},
  \bibinfo {author} {\bibfnamefont {J.-L.}\ \bibnamefont {Ville}}, \bibinfo
  {author} {\bibfnamefont {S.}~\bibnamefont {Nascimbene}}, \bibinfo {author}
  {\bibfnamefont {J.}~\bibnamefont {Beugnon}}, \ and\ \bibinfo {author}
  {\bibfnamefont {J.}~\bibnamefont {Dalibard}},\ }\bibfield  {title} {\bibinfo
  {title} {Dynamical symmetry and breathers in a two-dimensional Bose gas},\
  }\href {\doibase/10.1103/PhysRevX.9.021035} {\bibfield  {journal} {\bibinfo
  {journal} {Phys. Rev. X}\ }\textbf {\bibinfo {volume} {9}},\ \bibinfo {pages}
  {021035} (\bibinfo {year} {2019})}\BibitemShut {NoStop}%
\bibitem [{\citenamefont {Calogero}(1971)}]{Calogero1971}%
  \BibitemOpen
  \bibfield  {author} {\bibinfo {author} {\bibfnamefont {F.}~\bibnamefont
  {Calogero}},\ }\bibfield  {title} {\bibinfo {title} {Solution of the
  one-dimensional $N$-body problems with quadratic and/or inversely quadratic
  pair potentials},\ }\href {\doibase/10.1063/1.1665604} {\bibfield  {journal}
  {\bibinfo  {journal} {Journal of Mathematical Physics}\ }\textbf {\bibinfo
  {volume} {12}},\ \bibinfo {pages} {419} (\bibinfo {year} {1971})}\BibitemShut
  {NoStop}%
\bibitem [{\citenamefont {Sutherland}(1971)}]{Sutherland1971}%
  \BibitemOpen
  \bibfield  {author} {\bibinfo {author} {\bibfnamefont {B.}~\bibnamefont
  {Sutherland}},\ }\bibfield  {title} {\bibinfo {title} {Quantum many-body
  problem in one dimension: Ground state},\ }\href {\doibase/10.1063/1.1665584}
  {\bibfield  {journal} {\bibinfo  {journal} {Journal of Mathematical Physics}\
  }\textbf {\bibinfo {volume} {12}},\ \bibinfo {pages} {246} (\bibinfo {year}
  {1971})}\BibitemShut {NoStop}%
\bibitem [{\citenamefont {del Campo}(2016)}]{Campo2016}%
  \BibitemOpen
  \bibfield  {author} {\bibinfo {author} {\bibfnamefont {A.}~\bibnamefont {del
  Campo}},\ }\bibfield  {title} {\bibinfo {title} {Exact quantum decay of an
  interacting many-particle system: the Calogero{\textendash}Sutherland
  model},\ }\href {\doibase/10.1088/1367-2630/18/1/015014} {\bibfield
  {journal} {\bibinfo  {journal} {New Journal of Physics}\ }\textbf {\bibinfo
  {volume} {18}},\ \bibinfo {pages} {015014} (\bibinfo {year}
  {2016})}\BibitemShut {NoStop}%
\bibitem [{\citenamefont {Yurke}\ \emph {et~al.}(1986)\citenamefont {Yurke},
  \citenamefont {McCall},\ and\ \citenamefont {Klauder}}]{Yurke1986}%
  \BibitemOpen
  \bibfield  {author} {\bibinfo {author} {\bibfnamefont {B.}~\bibnamefont
  {Yurke}}, \bibinfo {author} {\bibfnamefont {S.~L.}\ \bibnamefont {McCall}}, \
  and\ \bibinfo {author} {\bibfnamefont {J.~R.}\ \bibnamefont {Klauder}},\
  }\bibfield  {title} {\bibinfo {title} {SU(2) and SU(1,1) interferometers},\
  }\href {\doibase/10.1103/PhysRevA.33.4033} {\bibfield  {journal} {\bibinfo
  {journal} {Phys. Rev. A}\ }\textbf {\bibinfo {volume} {33}},\ \bibinfo
  {pages} {4033} (\bibinfo {year} {1986})}\BibitemShut {NoStop}%
\bibitem [{\citenamefont {Linnemann}\ \emph {et~al.}(2016)\citenamefont
  {Linnemann}, \citenamefont {Strobel}, \citenamefont {Muessel}, \citenamefont
  {Schulz}, \citenamefont {Lewis-Swan}, \citenamefont {Kheruntsyan},\ and\
  \citenamefont {Oberthaler}}]{Linnemann2016}%
  \BibitemOpen
  \bibfield  {author} {\bibinfo {author} {\bibfnamefont {D.}~\bibnamefont
  {Linnemann}}, \bibinfo {author} {\bibfnamefont {H.}~\bibnamefont {Strobel}},
  \bibinfo {author} {\bibfnamefont {W.}~\bibnamefont {Muessel}}, \bibinfo
  {author} {\bibfnamefont {J.}~\bibnamefont {Schulz}}, \bibinfo {author}
  {\bibfnamefont {R.~J.}\ \bibnamefont {Lewis-Swan}}, \bibinfo {author}
  {\bibfnamefont {K.~V.}\ \bibnamefont {Kheruntsyan}}, \ and\ \bibinfo {author}
  {\bibfnamefont {M.~K.}\ \bibnamefont {Oberthaler}},\ }\bibfield  {title}
  {\bibinfo {title} {Quantum-enhanced sensing based on time reversal of
  nonlinear dynamics},\ }\href {\doibase/10.1103/PhysRevLett.117.013001}
  {\bibfield  {journal} {\bibinfo  {journal} {Phys. Rev. Lett.}\ }\textbf
  {\bibinfo {volume} {117}},\ \bibinfo {pages} {013001} (\bibinfo {year}
  {2016})}\BibitemShut {NoStop}%
\bibitem [{\citenamefont {Hu}\ \emph {et~al.}(2019)\citenamefont {Hu},
  \citenamefont {Feng}, \citenamefont {Zhang},\ and\ \citenamefont
  {Chin}}]{Hu2019}%
  \BibitemOpen
  \bibfield  {author} {\bibinfo {author} {\bibfnamefont {J.}~\bibnamefont
  {Hu}}, \bibinfo {author} {\bibfnamefont {L.}~\bibnamefont {Feng}}, \bibinfo
  {author} {\bibfnamefont {Z.}~\bibnamefont {Zhang}}, \ and\ \bibinfo {author}
  {\bibfnamefont {C.}~\bibnamefont {Chin}},\ }\bibfield  {title} {\bibinfo
  {title} {Quantum simulation of Unruh radiation},\ }\href
  {\doibase/10.1038/s41567-019-0537-1} {\bibfield  {journal} {\bibinfo
  {journal} {Nature Physics}\ }\textbf {\bibinfo {volume} {15}},\ \bibinfo
  {pages} {785} (\bibinfo {year} {2019})}\BibitemShut {NoStop}%
\bibitem [{\citenamefont {Kim}\ \emph {et~al.}(2021)\citenamefont {Kim},
  \citenamefont {Hur}, \citenamefont {Huh}, \citenamefont {Choi},\ and\
  \citenamefont {Choi}}]{Choi2021}%
  \BibitemOpen
  \bibfield  {author} {\bibinfo {author} {\bibfnamefont {K.}~\bibnamefont
  {Kim}}, \bibinfo {author} {\bibfnamefont {J.}~\bibnamefont {Hur}}, \bibinfo
  {author} {\bibfnamefont {S.J.}~\bibnamefont {Huh}}, \bibinfo {author}
  {\bibfnamefont {S.}~\bibnamefont {Choi}}, \ and\ \bibinfo {author}
  {\bibfnamefont {J.-Y.}\ \bibnamefont {Choi}},\ }\bibfield  {title} {\bibinfo
  {title} {Emission of spin-correlated matter-wave jets from spinor
  Bose-Einstein condensates},\ }\href {\doibase/10.1103/PhysRevLett.127.043401}
  {\bibfield  {journal} {\bibinfo  {journal} {Phys. Rev. Lett.}\ }\textbf
  {\bibinfo {volume} {127}},\ \bibinfo {pages} {043401} (\bibinfo {year}
  {2021})}\BibitemShut {NoStop}%
\bibitem [{\citenamefont {Guan}\ \emph {et~al.}(2021)\citenamefont {Guan},
  \citenamefont {Biedermann}, \citenamefont {Schwettmann},\ and\ \citenamefont
  {Lewis-Swan}}]{Guan2021}%
  \BibitemOpen
  \bibfield  {author} {\bibinfo {author} {\bibfnamefont {Q.}~\bibnamefont
  {Guan}}, \bibinfo {author} {\bibfnamefont {G.~W.}\ \bibnamefont
  {Biedermann}}, \bibinfo {author} {\bibfnamefont {A.}~\bibnamefont
  {Schwettmann}}, \ and\ \bibinfo {author} {\bibfnamefont {R.~J.}\ \bibnamefont
  {Lewis-Swan}},\ }\bibfield  {title} {\bibinfo {title} {Tailored generation of
  quantum states in an entangled spinor interferometer to overcome detection
  noise},\ }\href {\doibase/10.1103/PhysRevA.104.042415} {\bibfield  {journal}
  {\bibinfo  {journal} {Phys. Rev. A}\ }\textbf {\bibinfo {volume} {104}},\
  \bibinfo {pages} {042415} (\bibinfo {year} {2021})}\BibitemShut {NoStop}%
\bibitem [{\citenamefont {Mukherjee}\ \emph {et~al.}(2022)\citenamefont
  {Mukherjee}, \citenamefont {Shaffer}, \citenamefont {Patel}, \citenamefont
  {Yan}, \citenamefont {Wilson}, \citenamefont {Cr{\'{e}}pel}, \citenamefont
  {Fletcher},\ and\ \citenamefont {Zwierlein}}]{Mukherjee2022}%
  \BibitemOpen
  \bibfield  {author} {\bibinfo {author} {\bibfnamefont {B.}~\bibnamefont
  {Mukherjee}}, \bibinfo {author} {\bibfnamefont {A.}~\bibnamefont {Shaffer}},
  \bibinfo {author} {\bibfnamefont {P.~B.}\ \bibnamefont {Patel}}, \bibinfo
  {author} {\bibfnamefont {Z.}~\bibnamefont {Yan}}, \bibinfo {author}
  {\bibfnamefont {C.~C.}\ \bibnamefont {Wilson}}, \bibinfo {author}
  {\bibfnamefont {V.}~\bibnamefont {Cr{\'{e}}pel}}, \bibinfo {author}
  {\bibfnamefont {R.~J.}\ \bibnamefont {Fletcher}}, \ and\ \bibinfo {author}
  {\bibfnamefont {M.}~\bibnamefont {Zwierlein}},\ }\bibfield  {title} {\bibinfo
  {title} {Crystallization of bosonic quantum Hall states in a rotating quantum
  gas},\ }\href {\doibase/10.1038/s41586-021-04170-2} {\bibfield  {journal}
  {\bibinfo  {journal} {Nature}\ }\textbf {\bibinfo {volume} {601}},\ \bibinfo
  {pages} {58} (\bibinfo {year} {2022})}\BibitemShut {NoStop}%
\bibitem [{\citenamefont {Donley}\ \emph {et~al.}(2001)\citenamefont {Donley},
  \citenamefont {Claussen}, \citenamefont {Cornish}, \citenamefont {Roberts},
  \citenamefont {Cornell},\ and\ \citenamefont {Wieman}}]{Donley2001}%
  \BibitemOpen
  \bibfield  {author} {\bibinfo {author} {\bibfnamefont {E.~A.}\ \bibnamefont
  {Donley}}, \bibinfo {author} {\bibfnamefont {N.~R.}\ \bibnamefont
  {Claussen}}, \bibinfo {author} {\bibfnamefont {S.~L.}\ \bibnamefont
  {Cornish}}, \bibinfo {author} {\bibfnamefont {J.~L.}\ \bibnamefont
  {Roberts}}, \bibinfo {author} {\bibfnamefont {E.~A.}\ \bibnamefont
  {Cornell}}, \ and\ \bibinfo {author} {\bibfnamefont {C.~E.}\ \bibnamefont
  {Wieman}},\ }\bibfield  {title} {\bibinfo {title} {Dynamics of collapsing and
  exploding Bose{\textendash}Einstein condensates},\ }\href
  {\doibase/10.1038/35085500} {\bibfield  {journal} {\bibinfo  {journal}
  {Nature}\ }\textbf {\bibinfo {volume} {412}},\ \bibinfo {pages} {295}
  (\bibinfo {year} {2001})}\BibitemShut {NoStop}%
\bibitem [{\citenamefont {Parker}\ \emph {et~al.}(2013)\citenamefont {Parker},
  \citenamefont {Ha},\ and\ \citenamefont {Chin}}]{Parker2013}%
  \BibitemOpen
  \bibfield  {author} {\bibinfo {author} {\bibfnamefont {C.~V.}\ \bibnamefont
  {Parker}}, \bibinfo {author} {\bibfnamefont {L.-C.}\ \bibnamefont {Ha}}, \
  and\ \bibinfo {author} {\bibfnamefont {C.}~\bibnamefont {Chin}},\ }\bibfield
  {title} {\bibinfo {title} {Direct observation of effective ferromagnetic
  domains of cold atoms in a shaken optical lattice},\ }\href
  {\doibase/10.1038/nphys2789} {\bibfield  {journal} {\bibinfo  {journal}
  {Nature Physics}\ }\textbf {\bibinfo {volume} {9}},\ \bibinfo {pages} {769}
  (\bibinfo {year} {2013})}\BibitemShut {NoStop}%
\bibitem [{\citenamefont {Nguyen}\ \emph {et~al.}(2017)\citenamefont {Nguyen},
  \citenamefont {Luo},\ and\ \citenamefont {Hulet}}]{Nguyen2017}%
  \BibitemOpen
  \bibfield  {author} {\bibinfo {author} {\bibfnamefont {J.~H.~V.}\
  \bibnamefont {Nguyen}}, \bibinfo {author} {\bibfnamefont {D.}~\bibnamefont
  {Luo}}, \ and\ \bibinfo {author} {\bibfnamefont {R.~G.}\ \bibnamefont
  {Hulet}},\ }\bibfield  {title} {\bibinfo {title} {Formation of matter-wave
  soliton trains by modulational instability},\ }\href
  {\doibase/10.1126/science.aal3220} {\bibfield  {journal} {\bibinfo  {journal}
  {Science}\ }\textbf {\bibinfo {volume} {356}},\ \bibinfo {pages} {422}
  (\bibinfo {year} {2017})}\BibitemShut {NoStop}%
\bibitem [{\citenamefont {Wintersperger}\ \emph {et~al.}(2020)\citenamefont
  {Wintersperger}, \citenamefont {Bukov}, \citenamefont {N\"ager},
  \citenamefont {Lellouch}, \citenamefont {Demler}, \citenamefont {Schneider},
  \citenamefont {Bloch}, \citenamefont {Goldman},\ and\ \citenamefont
  {Aidelsburger}}]{Bloch2020}%
  \BibitemOpen
  \bibfield  {author} {\bibinfo {author} {\bibfnamefont {K.}~\bibnamefont
  {Wintersperger}}, \bibinfo {author} {\bibfnamefont {M.}~\bibnamefont
  {Bukov}}, \bibinfo {author} {\bibfnamefont {J.}~\bibnamefont {N\"ager}},
  \bibinfo {author} {\bibfnamefont {S.}~\bibnamefont {Lellouch}}, \bibinfo
  {author} {\bibfnamefont {E.}~\bibnamefont {Demler}}, \bibinfo {author}
  {\bibfnamefont {U.}~\bibnamefont {Schneider}}, \bibinfo {author}
  {\bibfnamefont {I.}~\bibnamefont {Bloch}}, \bibinfo {author} {\bibfnamefont
  {N.}~\bibnamefont {Goldman}}, \ and\ \bibinfo {author} {\bibfnamefont
  {M.}~\bibnamefont {Aidelsburger}},\ }\bibfield  {title} {\bibinfo {title}
  {Parametric instabilities of interacting bosons in periodically driven 1D
  optical lattices},\ }\href {\doibase/10.1103/PhysRevX.10.011030} {\bibfield
  {journal} {\bibinfo  {journal} {Phys. Rev. X}\ }\textbf {\bibinfo {volume}
  {10}},\ \bibinfo {pages} {011030} (\bibinfo {year} {2020})}\BibitemShut
  {NoStop}%
\bibitem [{\citenamefont {Chen}\ and\ \citenamefont {Hung}(2020)}]{Hung2020}%
  \BibitemOpen
  \bibfield  {author} {\bibinfo {author} {\bibfnamefont {C.-A.}\ \bibnamefont
  {Chen}}\ and\ \bibinfo {author} {\bibfnamefont {C.-L.}\ \bibnamefont
  {Hung}},\ }\bibfield  {title} {\bibinfo {title} {Observation of universal
  quench dynamics and Townes soliton formation from modulational instability in
  two-dimensional Bose gases},\ }\href
  {\doibase/10.1103/PhysRevLett.125.250401} {\bibfield  {journal} {\bibinfo
  {journal} {Phys. Rev. Lett.}\ }\textbf {\bibinfo {volume} {125}},\ \bibinfo
  {pages} {250401} (\bibinfo {year} {2020})}\BibitemShut {NoStop}%
\bibitem [{\citenamefont {Lapierre}\ \emph {et~al.}(2020)\citenamefont
  {Lapierre}, \citenamefont {Choo}, \citenamefont {Tauber}, \citenamefont
  {Tiwari}, \citenamefont {Neupert},\ and\ \citenamefont
  {Chitra}}]{Lapierre2020}%
  \BibitemOpen
  \bibfield  {author} {\bibinfo {author} {\bibfnamefont {B.}~\bibnamefont
  {Lapierre}}, \bibinfo {author} {\bibfnamefont {K.}~\bibnamefont {Choo}},
  \bibinfo {author} {\bibfnamefont {C.}~\bibnamefont {Tauber}}, \bibinfo
  {author} {\bibfnamefont {A.}~\bibnamefont {Tiwari}}, \bibinfo {author}
  {\bibfnamefont {T.}~\bibnamefont {Neupert}}, \ and\ \bibinfo {author}
  {\bibfnamefont {R.}~\bibnamefont {Chitra}},\ }\bibfield  {title} {\bibinfo
  {title} {Emergent black hole dynamics in critical Floquet systems},\ }\href
  {\doibase/10.1103/PhysRevResearch.2.023085} {\bibfield  {journal} {\bibinfo
  {journal} {Phys. Rev. Research}\ }\textbf {\bibinfo {volume} {2}},\ \bibinfo
  {pages} {023085} (\bibinfo {year} {2020})}\BibitemShut {NoStop}%
\bibitem [{\citenamefont {Lapierre}\ and\ \citenamefont
  {Moosavi}(2021)}]{Moosavi2021}%
  \BibitemOpen
  \bibfield  {author} {\bibinfo {author} {\bibfnamefont {B.}~\bibnamefont
  {Lapierre}}\ and\ \bibinfo {author} {\bibfnamefont {P.}~\bibnamefont
  {Moosavi}},\ }\bibfield  {title} {\bibinfo {title} {Geometric approach to
  inhomogeneous Floquet systems},\ }\href
  {\doibase/10.1103/PhysRevB.103.224303} {\bibfield  {journal} {\bibinfo
  {journal} {Phys. Rev. B}\ }\textbf {\bibinfo {volume} {103}},\ \bibinfo
  {pages} {224303} (\bibinfo {year} {2021})}\BibitemShut {NoStop}%
\bibitem [{\citenamefont {Fan}\ \emph {et~al.}(2021)\citenamefont {Fan},
  \citenamefont {Gu}, \citenamefont {Vishwanath},\ and\ \citenamefont
  {Wen}}]{Wen2021}%
  \BibitemOpen
  \bibfield  {author} {\bibinfo {author} {\bibfnamefont {R.}~\bibnamefont
  {Fan}}, \bibinfo {author} {\bibfnamefont {Y.}~\bibnamefont {Gu}}, \bibinfo
  {author} {\bibfnamefont {A.}~\bibnamefont {Vishwanath}}, \ and\ \bibinfo
  {author} {\bibfnamefont {X.}~\bibnamefont {Wen}},\ }\bibfield  {title}
  {\bibinfo {title} {{Floquet conformal field theories with generally deformed
  Hamiltonians}},\ }\href {\doibase/10.21468/SciPostPhys.10.2.049} {\bibfield
  {journal} {\bibinfo  {journal} {SciPost Phys.}\ }\textbf {\bibinfo {volume}
  {10}},\ \bibinfo {pages} {49} (\bibinfo {year} {2021})}\BibitemShut {NoStop}%
\bibitem [{\citenamefont {Wiersig}(2014)}]{Wiersig2014}%
  \BibitemOpen
  \bibfield  {author} {\bibinfo {author} {\bibfnamefont {J.}~\bibnamefont
  {Wiersig}},\ }\bibfield  {title} {\bibinfo {title} {Enhancing the sensitivity
  of frequency and energy splitting detection by using exceptional points:
  Application to microcavity sensors for single-particle detection},\ }\href
  {\doibase/10.1103/PhysRevLett.112.203901} {\bibfield  {journal} {\bibinfo
  {journal} {Phys. Rev. Lett.}\ }\textbf {\bibinfo {volume} {112}},\ \bibinfo
  {pages} {203901} (\bibinfo {year} {2014})}\BibitemShut {NoStop}%
\bibitem [{\citenamefont {Wiersig}(2016)}]{Wiersig2016}%
  \BibitemOpen
  \bibfield  {author} {\bibinfo {author} {\bibfnamefont {J.}~\bibnamefont
  {Wiersig}},\ }\bibfield  {title} {\bibinfo {title} {Sensors operating at
  exceptional points: General theory},\ }\href
  {\doibase/10.1103/PhysRevA.93.033809} {\bibfield  {journal} {\bibinfo
  {journal} {Phys. Rev. A}\ }\textbf {\bibinfo {volume} {93}},\ \bibinfo
  {pages} {033809} (\bibinfo {year} {2016})}\BibitemShut {NoStop}%
\bibitem [{\citenamefont {Liu}\ \emph {et~al.}(2016)\citenamefont {Liu},
  \citenamefont {Zhang}, \citenamefont {\"Ozdemir}, \citenamefont {Peng},
  \citenamefont {Jing}, \citenamefont {L\"u}, \citenamefont {Li}, \citenamefont
  {Yang}, \citenamefont {Nori},\ and\ \citenamefont {Liu}}]{Liu2016}%
  \BibitemOpen
  \bibfield  {author} {\bibinfo {author} {\bibfnamefont {Z.-P.}\ \bibnamefont
  {Liu}}, \bibinfo {author} {\bibfnamefont {J.}~\bibnamefont {Zhang}}, \bibinfo
  {author} {\bibfnamefont {{\c{S}}.~K.}\ \bibnamefont {\"Ozdemir}}, \bibinfo
  {author} {\bibfnamefont {B.}~\bibnamefont {Peng}}, \bibinfo {author}
  {\bibfnamefont {H.}~\bibnamefont {Jing}}, \bibinfo {author} {\bibfnamefont
  {X.-Y.}\ \bibnamefont {L\"u}}, \bibinfo {author} {\bibfnamefont {C.-W.}\
  \bibnamefont {Li}}, \bibinfo {author} {\bibfnamefont {L.}~\bibnamefont
  {Yang}}, \bibinfo {author} {\bibfnamefont {F.}~\bibnamefont {Nori}}, \ and\
  \bibinfo {author} {\bibfnamefont {Y.-x.}\ \bibnamefont {Liu}},\ }\bibfield
  {title} {\bibinfo {title} {Metrology with $\mathcal{PT}$-symmetric cavities:
  Enhanced sensitivity near the $\mathcal{PT}$-phase transition},\ }\href
  {\doibase/10.1103/PhysRevLett.117.110802} {\bibfield  {journal} {\bibinfo
  {journal} {Phys. Rev. Lett.}\ }\textbf {\bibinfo {volume} {117}},\ \bibinfo
  {pages} {110802} (\bibinfo {year} {2016})}\BibitemShut {NoStop}%
\bibitem [{\citenamefont {Hodaei}\ \emph {et~al.}(2017)\citenamefont {Hodaei},
  \citenamefont {Hassan}, \citenamefont {Wittek}, \citenamefont
  {Garcia-Gracia}, \citenamefont {El-Ganainy}, \citenamefont
  {Christodoulides},\ and\ \citenamefont {Khajavikhan}}]{Hodaei2017}%
  \BibitemOpen
  \bibfield  {author} {\bibinfo {author} {\bibfnamefont {H.}~\bibnamefont
  {Hodaei}}, \bibinfo {author} {\bibfnamefont {A.~U.}\ \bibnamefont {Hassan}},
  \bibinfo {author} {\bibfnamefont {S.}~\bibnamefont {Wittek}}, \bibinfo
  {author} {\bibfnamefont {H.}~\bibnamefont {Garcia-Gracia}}, \bibinfo {author}
  {\bibfnamefont {R.}~\bibnamefont {El-Ganainy}}, \bibinfo {author}
  {\bibfnamefont {D.~N.}\ \bibnamefont {Christodoulides}}, \ and\ \bibinfo
  {author} {\bibfnamefont {M.}~\bibnamefont {Khajavikhan}},\ }\bibfield
  {title} {\bibinfo {title} {Enhanced sensitivity at higher-order exceptional
  points},\ }\href {\doibase/10.1038/nature23280} {\bibfield  {journal}
  {\bibinfo  {journal} {Nature}\ }\textbf {\bibinfo {volume} {548}},\ \bibinfo
  {pages} {187} (\bibinfo {year} {2017})}\BibitemShut {NoStop}%
\bibitem [{\citenamefont {Chen}\ \emph {et~al.}(2017)\citenamefont {Chen},
  \citenamefont {\"{O}zdemir}, \citenamefont {Zhao}, \citenamefont {Wiersig},\
  and\ \citenamefont {Yang}}]{Chen2017}%
  \BibitemOpen
  \bibfield  {author} {\bibinfo {author} {\bibfnamefont {W.}~\bibnamefont
  {Chen}}, \bibinfo {author} {\bibfnamefont {{\c{S}}.~K.}\ \bibnamefont
  {\"{O}zdemir}}, \bibinfo {author} {\bibfnamefont {G.}~\bibnamefont {Zhao}},
  \bibinfo {author} {\bibfnamefont {J.}~\bibnamefont {Wiersig}}, \ and\
  \bibinfo {author} {\bibfnamefont {L.}~\bibnamefont {Yang}},\ }\bibfield
  {title} {\bibinfo {title} {Exceptional points enhance sensing in an optical
  microcavity},\ }\href {\doibase/10.1038/nature23281} {\bibfield  {journal}
  {\bibinfo  {journal} {Nature}\ }\textbf {\bibinfo {volume} {548}},\ \bibinfo
  {pages} {192} (\bibinfo {year} {2017})}\BibitemShut {NoStop}%
\bibitem [{\citenamefont {Lau}\ and\ \citenamefont {Clerk}(2018)}]{Lau2018}%
  \BibitemOpen
  \bibfield  {author} {\bibinfo {author} {\bibfnamefont {H.-K.}\ \bibnamefont
  {Lau}}\ and\ \bibinfo {author} {\bibfnamefont {A.~A.}\ \bibnamefont
  {Clerk}},\ }\bibfield  {title} {\bibinfo {title} {Fundamental limits and
  non-reciprocal approaches in non-Hermitian quantum sensing},\ }\href
  {\doibase/10.1038/s41467-018-06477-7} {\bibfield  {journal} {\bibinfo
  {journal} {Nature Communications}\ }\textbf {\bibinfo {volume} {9}} (\bibinfo
  {year} {2018}),\ 10.1038/s41467-018-06477-7}\BibitemShut {NoStop}%
\bibitem [{\citenamefont {Zhang}\ \emph {et~al.}(2019)\citenamefont {Zhang},
  \citenamefont {Sweeney}, \citenamefont {Hsu}, \citenamefont {Yang},
  \citenamefont {Stone},\ and\ \citenamefont {Jiang}}]{Zhang2019}%
  \BibitemOpen
  \bibfield  {author} {\bibinfo {author} {\bibfnamefont {M.}~\bibnamefont
  {Zhang}}, \bibinfo {author} {\bibfnamefont {W.}~\bibnamefont {Sweeney}},
  \bibinfo {author} {\bibfnamefont {C.~W.}\ \bibnamefont {Hsu}}, \bibinfo
  {author} {\bibfnamefont {L.}~\bibnamefont {Yang}}, \bibinfo {author}
  {\bibfnamefont {A.~D.}\ \bibnamefont {Stone}}, \ and\ \bibinfo {author}
  {\bibfnamefont {L.}~\bibnamefont {Jiang}},\ }\bibfield  {title} {\bibinfo
  {title} {Quantum noise theory of exceptional point amplifying sensors},\
  }\href {\doibase/10.1103/PhysRevLett.123.180501} {\bibfield  {journal}
  {\bibinfo  {journal} {Phys. Rev. Lett.}\ }\textbf {\bibinfo {volume} {123}},\
  \bibinfo {pages} {180501} (\bibinfo {year} {2019})}\BibitemShut {NoStop}%
\bibitem [{\citenamefont {Hokmabadi}\ \emph {et~al.}(2019)\citenamefont
  {Hokmabadi}, \citenamefont {Schumer}, \citenamefont {Christodoulides},\ and\
  \citenamefont {Khajavikhan}}]{Hokmabadi2019}%
  \BibitemOpen
  \bibfield  {author} {\bibinfo {author} {\bibfnamefont {M.~P.}\ \bibnamefont
  {Hokmabadi}}, \bibinfo {author} {\bibfnamefont {A.}~\bibnamefont {Schumer}},
  \bibinfo {author} {\bibfnamefont {D.~N.}\ \bibnamefont {Christodoulides}}, \
  and\ \bibinfo {author} {\bibfnamefont {M.}~\bibnamefont {Khajavikhan}},\
  }\bibfield  {title} {\bibinfo {title} {Non-Hermitian ring~laser gyroscopes
  with enhanced Sagnac sensitivity},\ }\href
  {\doibase/10.1038/s41586-019-1780-4} {\bibfield  {journal} {\bibinfo
  {journal} {Nature}\ }\textbf {\bibinfo {volume} {576}},\ \bibinfo {pages}
  {70} (\bibinfo {year} {2019})}\BibitemShut {NoStop}%
\bibitem [{\citenamefont {Lai}\ \emph {et~al.}(2019)\citenamefont {Lai},
  \citenamefont {Lu}, \citenamefont {Suh}, \citenamefont {Yuan},\ and\
  \citenamefont {Vahala}}]{Lai2019}%
  \BibitemOpen
  \bibfield  {author} {\bibinfo {author} {\bibfnamefont {Y.-H.}\ \bibnamefont
  {Lai}}, \bibinfo {author} {\bibfnamefont {Y.-K.}\ \bibnamefont {Lu}},
  \bibinfo {author} {\bibfnamefont {M.-G.}\ \bibnamefont {Suh}}, \bibinfo
  {author} {\bibfnamefont {Z.}~\bibnamefont {Yuan}}, \ and\ \bibinfo {author}
  {\bibfnamefont {K.}~\bibnamefont {Vahala}},\ }\bibfield  {title} {\bibinfo
  {title} {Observation of the exceptional-point-enhanced Sagnac effect},\
  }\href {\doibase/10.1038/s41586-019-1777-z} {\bibfield  {journal} {\bibinfo
  {journal} {Nature}\ }\textbf {\bibinfo {volume} {576}},\ \bibinfo {pages}
  {65} (\bibinfo {year} {2019})}\BibitemShut {NoStop}%
\bibitem [{\citenamefont {Torrontegui}\ \emph {et~al.}(2013)\citenamefont
  {Torrontegui}, \citenamefont {Ib{\'{a}}{\~{n}}ez}, \citenamefont
  {Mart{\'{\i}}nez-Garaot}, \citenamefont {Modugno}, \citenamefont {del Campo},
  \citenamefont {Gu{\'{e}}ry-Odelin}, \citenamefont {Ruschhaupt}, \citenamefont
  {Chen},\ and\ \citenamefont {Muga}}]{Torrontegui2013}%
  \BibitemOpen
  \bibfield  {author} {\bibinfo {author} {\bibfnamefont {E.}~\bibnamefont
  {Torrontegui}}, \bibinfo {author} {\bibfnamefont {S.}~\bibnamefont
  {Ib{\'{a}}{\~{n}}ez}}, \bibinfo {author} {\bibfnamefont {S.}~\bibnamefont
  {Mart{\'{\i}}nez-Garaot}}, \bibinfo {author} {\bibfnamefont {M.}~\bibnamefont
  {Modugno}}, \bibinfo {author} {\bibfnamefont {A.}~\bibnamefont {del Campo}},
  \bibinfo {author} {\bibfnamefont {D.}~\bibnamefont {Gu{\'{e}}ry-Odelin}},
  \bibinfo {author} {\bibfnamefont {A.}~\bibnamefont {Ruschhaupt}}, \bibinfo
  {author} {\bibfnamefont {X.}~\bibnamefont {Chen}}, \ and\ \bibinfo {author}
  {\bibfnamefont {J.~G.}\ \bibnamefont {Muga}},\ }\bibfield  {title} {\bibinfo
  {title} {Shortcuts to adiabaticity},\ }in\ \href
  {\doibase/10.1016/b978-0-12-408090-4.00002-5} { {\bibinfo {booktitle}
  {Advances In Atomic, Molecular, and Optical Physics}}}\ (\bibinfo
  {publisher} {Elsevier},\ \bibinfo {year} {2013})\ pp.\ \bibinfo {pages}
  {117--169}\BibitemShut {NoStop}%
\bibitem [{\citenamefont {Reiserer}\ \emph {et~al.}(2013)\citenamefont
  {Reiserer}, \citenamefont {Ritter},\ and\ \citenamefont
  {Rempe}}]{Reiserer2013}%
  \BibitemOpen
  \bibfield  {author} {\bibinfo {author} {\bibfnamefont {A.}~\bibnamefont
  {Reiserer}}, \bibinfo {author} {\bibfnamefont {S.}~\bibnamefont {Ritter}}, \
  and\ \bibinfo {author} {\bibfnamefont {G.}~\bibnamefont {Rempe}},\ }\bibfield
   {title} {\bibinfo {title} {Nondestructive detection of an optical photon},\
  }\href {\doibase/10.1126/science.1246164} {\bibfield  {journal} {\bibinfo
  {journal} {Science}\ }\textbf {\bibinfo {volume} {342}},\ \bibinfo {pages}
  {1349} (\bibinfo {year} {2013})}\BibitemShut {NoStop}%
\bibitem [{\citenamefont {Cetina}\ \emph {et~al.}(2016)\citenamefont {Cetina},
  \citenamefont {Jag}, \citenamefont {Lous}, \citenamefont {Fritsche},
  \citenamefont {Walraven}, \citenamefont {Grimm}, \citenamefont {Levinsen},
  \citenamefont {Parish}, \citenamefont {Schmidt}, \citenamefont {Knap},\ and\
  \citenamefont {Demler}}]{Cetina2016}%
  \BibitemOpen
  \bibfield  {author} {\bibinfo {author} {\bibfnamefont {M.}~\bibnamefont
  {Cetina}}, \bibinfo {author} {\bibfnamefont {M.}~\bibnamefont {Jag}},
  \bibinfo {author} {\bibfnamefont {R.~S.}\ \bibnamefont {Lous}}, \bibinfo
  {author} {\bibfnamefont {I.}~\bibnamefont {Fritsche}}, \bibinfo {author}
  {\bibfnamefont {J.~T.~M.}\ \bibnamefont {Walraven}}, \bibinfo {author}
  {\bibfnamefont {R.}~\bibnamefont {Grimm}}, \bibinfo {author} {\bibfnamefont
  {J.}~\bibnamefont {Levinsen}}, \bibinfo {author} {\bibfnamefont {M.~M.}\
  \bibnamefont {Parish}}, \bibinfo {author} {\bibfnamefont {R.}~\bibnamefont
  {Schmidt}}, \bibinfo {author} {\bibfnamefont {M.}~\bibnamefont {Knap}}, \
  and\ \bibinfo {author} {\bibfnamefont {E.}~\bibnamefont {Demler}},\
  }\bibfield  {title} {\bibinfo {title} {Ultrafast many-body interferometry of
  impurities coupled to a Fermi sea},\ }\href
  {\doibase/10.1126/science.aaf5134} {\bibfield  {journal} {\bibinfo  {journal}
  {Science}\ }\textbf {\bibinfo {volume} {354}},\ \bibinfo {pages} {96}
  (\bibinfo {year} {2016})}\BibitemShut {NoStop}%
\bibitem [{\citenamefont {Yan}\ and\ \citenamefont {Zhou}(2021)}]{Qi2021}%
  \BibitemOpen
  \bibfield  {author} {\bibinfo {author} {\bibfnamefont {Y.}~\bibnamefont
  {Yan}}\ and\ \bibinfo {author} {\bibfnamefont {Q.}~\bibnamefont {Zhou}},\
  }\href {\doibase/10.48550/ARXIV.2106.15726} {\bibinfo {title} {Manipulating
  anyons in quantum Hall droplets of light using dissipations},\ arXiv: 2106.15726} (\bibinfo
  {year} {2021})\BibitemShut {NoStop}%
  \bibitem [{\citenamefont {Heyl}\ \emph {et~al.}(2013)\citenamefont {Heyl},
  \citenamefont {Polkovnikov},\ and\ \citenamefont {Kehrein}}]{Heyl2013}%
  \BibitemOpen
  \bibfield  {author} {\bibinfo {author} {\bibfnamefont {M.}~\bibnamefont
  {Heyl}}, \bibinfo {author} {\bibfnamefont {A.}~\bibnamefont {Polkovnikov}},\
  and\ \bibinfo {author} {\bibfnamefont {S.}~\bibnamefont {Kehrein}},\
  }\bibfield  {title} {\bibinfo {title} {Dynamical quantum phase transitions in
  the transverse-field Ising model},\ }\href
  {https://doi.org/10.1103/PhysRevLett.110.135704} {\bibfield  {journal}
  {\bibinfo  {journal} {Phys. Rev. Lett.}\ }\textbf {\bibinfo {volume} {110}},\
  \bibinfo {pages} {135704} (\bibinfo {year} {2013})}\BibitemShut {NoStop}%
\end{thebibliography}

\begin{thebibliography}{7}%
\makeatletter
\providecommand \@ifxundefined [1]{%
 \@ifx{#1\undefined}
}%
\providecommand \@ifnum [1]{%
 \ifnum #1\expandafter \@firstoftwo
 \else \expandafter \@secondoftwo
 \fi
}%
\providecommand \@ifx [1]{%
 \ifx #1\expandafter \@firstoftwo
 \else \expandafter \@secondoftwo
 \fi
}%
\providecommand \natexlab [1]{#1}%
\providecommand \enquote  [1]{#1}%
\providecommand \bibnamefont  [1]{#1}%
\providecommand \bibfnamefont [1]{#1}%
\providecommand \citenamefont [1]{#1}%
\providecommand \href@noop [0]{\@secondoftwo}%
\providecommand \href [0]{\begingroup \@sanitize@url \@href}%
\providecommand \@href[1]{\@@startlink{#1}\@@href}%
\providecommand \@@href[1]{\endgroup#1\@@endlink}%
\providecommand \@sanitize@url [0]{\catcode `\\12\catcode `\$12\catcode
  `\&12\catcode `\#12\catcode `\^12\catcode `\_12\catcode `\%12\relax}%
\providecommand \@@startlink[1]{}%
\providecommand \@@endlink[0]{}%
\providecommand \url  [0]{\begingroup\@sanitize@url \@url }%
\providecommand \@url [1]{\endgroup\@href {#1}{\urlprefix }}%
\providecommand \urlprefix  [0]{URL }%
\providecommand \Eprint [0]{\href }%
\providecommand \doibase [0]{https://dx.doi.org}%
\providecommand \selectlanguage [0]{\@gobble}%
\providecommand \bibinfo  [0]{\@secondoftwo}%
\providecommand \bibfield  [0]{\@secondoftwo}%
\providecommand \translation [1]{[#1]}%
\providecommand \BibitemOpen [0]{}%
\providecommand \bibitemStop [0]{}%
\providecommand \bibitemNoStop [0]{.\EOS\space}%
\providecommand \EOS [0]{\spacefactor3000\relax}%
\providecommand \BibitemShut  [1]{\csname bibitem#1\endcsname}%
\let\auto@bib@innerbib\@empty
%</preamble>
\bibitem [{\citenamefont {Brown}\ and\ \citenamefont
  {Susskind}(2019)}]{SSusskind2019}%
  \BibitemOpen
  \bibfield  {author} {\bibinfo {author} {\bibfnamefont {A.~R.}\ \bibnamefont
  {Brown}}\ and\ \bibinfo {author} {\bibfnamefont {L.}~\bibnamefont
  {Susskind}},\ }\bibfield  {title} {\bibinfo {title} {Complexity geometry of a
  single qubit},\ }\href {\doibase/10.1103/PhysRevD.100.046020} {\bibfield
  {journal} {\bibinfo  {journal} {Phys. Rev. D}\ }\textbf {\bibinfo {volume}
  {100}},\ \bibinfo {pages} {046020} (\bibinfo {year} {2019})}\BibitemShut
  {NoStop}%
\bibitem [{\citenamefont {Bengtsson}\ and\ \citenamefont
  {Sandin}(2006)}]{SBengtsson2006}%
  \BibitemOpen
  \bibfield  {author} {\bibinfo {author} {\bibfnamefont {I.}~\bibnamefont
  {Bengtsson}}\ and\ \bibinfo {author} {\bibfnamefont {P.}~\bibnamefont
  {Sandin}},\ }\bibfield  {title} {\bibinfo {title} {Anti-de Sitter space,
  squashed and stretched},\ }\href {\doibase/10.1088/0264-9381/23/3/022}
  {\bibfield  {journal} {\bibinfo  {journal} {Classical and Quantum Gravity}\
  }\textbf {\bibinfo {volume} {23}},\ \bibinfo {pages} {971} (\bibinfo {year}
  {2006})}\BibitemShut {NoStop}%
\bibitem [{\citenamefont {'t~Hooft}(1974)}]{StHooft1974}%
  \BibitemOpen
  \bibfield  {author} {\bibinfo {author} {\bibfnamefont {G.}~\bibnamefont
  {'t~Hooft}},\ }\bibfield  {title} {\bibinfo {title} {Magnetic monopoles in
  unified theories},\ }\href@noop {} {\bibfield  {journal} {\bibinfo  {journal}
  {Nucl. Phys. B}\ }\textbf {\bibinfo {volume} {79}},\ \bibinfo {pages} {276}
  (\bibinfo {year} {1974})}\BibitemShut {NoStop}%
\bibitem [{\citenamefont {Polyakov}(1974)}]{SPolyakov1974}%
  \BibitemOpen
  \bibfield  {author} {\bibinfo {author} {\bibfnamefont {A.}~\bibnamefont
  {Polyakov}},\ }\bibfield  {title} {\bibinfo {title} {Particle spectrum in the
  quantum field theory},\ }\href@noop {} {\bibfield  {journal} {\bibinfo
  {journal} {Journal of Experimental and Theoretical Physics Letters (JETP
  Letters)}\ }\textbf {\bibinfo {volume} {20}},\ \bibinfo {pages} {194}
  (\bibinfo {year} {1974})}\BibitemShut {NoStop}%
\bibitem [{\citenamefont {Auzzi}\ \emph {et~al.}(2020)\citenamefont {Auzzi},
  \citenamefont {Baiguera}, \citenamefont {Legramandi}, \citenamefont
  {Nardelli}, \citenamefont {Roy},\ and\ \citenamefont {Zenoni}}]{SAuzzi2020}%
  \BibitemOpen
  \bibfield  {author} {\bibinfo {author} {\bibfnamefont {R.}~\bibnamefont
  {Auzzi}}, \bibinfo {author} {\bibfnamefont {S.}~\bibnamefont {Baiguera}},
  \bibinfo {author} {\bibfnamefont {A.}~\bibnamefont {Legramandi}}, \bibinfo
  {author} {\bibfnamefont {G.}~\bibnamefont {Nardelli}}, \bibinfo {author}
  {\bibfnamefont {P.}~\bibnamefont {Roy}}, \ and\ \bibinfo {author}
  {\bibfnamefont {N.}~\bibnamefont {Zenoni}},\ }\bibfield  {title} {\bibinfo
  {title} {On subregion action complexity in {AdS}3 and in the {BTZ} black
  hole},\ }\href {\doibase/10.1007/jhep01(2020)066} {\bibfield  {journal}
  {\bibinfo  {journal} {Journal of High Energy Physics}\ }\textbf {\bibinfo
  {volume} {2020}} (\bibinfo {year} {2020}),\
  10.1007/jhep01(2020)066}\BibitemShut {NoStop}%
\bibitem [{\citenamefont {Reynolds}\ and\ \citenamefont
  {Ross}(2018)}]{SReynolds2018}%
  \BibitemOpen
  \bibfield  {author} {\bibinfo {author} {\bibfnamefont {A.~P.}\ \bibnamefont
  {Reynolds}}\ and\ \bibinfo {author} {\bibfnamefont {S.~F.}\ \bibnamefont
  {Ross}},\ }\bibfield  {title} {\bibinfo {title} {Complexity of the {AdS}
  soliton},\ }\href {\doibase/10.1088/1361-6382/aab32d} {\bibfield  {journal}
  {\bibinfo  {journal} {Classical and Quantum Gravity}\ }\textbf {\bibinfo
  {volume} {35}},\ \bibinfo {pages} {095006} (\bibinfo {year}
  {2018})}\BibitemShut {NoStop}%
\bibitem [{\citenamefont {Carroll}(2019)}]{SCarroll2019}%
  \BibitemOpen
  \bibfield  {author} {\bibinfo {author} {\bibfnamefont {S.~M.}\ \bibnamefont
  {Carroll}},\ }\href@noop {} {{\bibinfo {title} {Spacetime and
  geometry}}}\ (\bibinfo  {publisher} {Cambridge University Press},\ \bibinfo
  {year} {2019})\BibitemShut {NoStop}%
\end{thebibliography}
\end{document}